\newcount\Stiloption

\ifnum\Stiloption=0
	\documentclass{iopart}
\else
	\documentclass[twoside]{article}
\fi

\usepackage[a4paper,left=3.5cm, right=3.5cm, top=21mm, bottom=29mm,includeheadfoot,marginparwidth=2.5cm]{geometry}

\ifnum\Stiloption=0
	\usepackage{iopams}
	\expandafter\let\csname equation*\endcsname=\relax
	\expandafter\let\csname endequation*\endcsname=\relax
\else
	\usepackage{amsmath}
\fi

\usepackage{amssymb}
\usepackage{amsbsy}
\usepackage{amsfonts}
\usepackage{graphicx}
\usepackage{verbatim}
\usepackage{enumitem}
\usepackage{rotating}
\usepackage{ifthen}

\usepackage{marginnote}
\usepackage[hang]{footmisc}

\usepackage[latin1]{inputenc}

\usepackage[T1]{fontenc}
\usepackage{lmodern}

\usepackage{fancyhdr}

\usepackage[english]{babel}

\usepackage{amsthm}
\usepackage{extpfeil}

\usepackage{xcolor}
\definecolor{refkey}{gray}{0.75}
\definecolor{labelkey}{gray}{0.5}
\definecolor{lightblue}{rgb}{0.6,0.6,1}
\definecolor{lightgray}{rgb}{0.97,0.97,0.97}
\definecolor{darkred}{rgb}{0.5,0.2,0.2}
\usepackage[colorlinks,pdfpagelabels,pdfstartview = FitH,bookmarksopen = true,bookmarksnumbered = true,linkcolor = red,plainpages = false,hypertexnames = false,citecolor = blue] {hyperref}

\numberwithin{equation}{section}

\renewcommand{\v}[1]{\ensuremath{\mathbf{#1}}} % bold type for vectors
\newcommand{\gv}[1]{\ensuremath{\mbox{\boldmath $ #1 $}}} % bold type for Greek letters
\newcommand{\grad}[1]{\gv{\nabla} #1} % for gradient
\renewcommand{\div}[1]{\gv{\nabla} \cdot #1} % for divergence
\newcommand{\rot}[1]{\gv{\nabla} \times #1} % for curl
\usepackage[squaren,gray,pstricks]{SIunits}

\theoremstyle{definition}

\theoremstyle{plain}

\newtheorem{prob}{Problem}

\newtheorem{remark}{Remark}

\usepackage{graphicx}
\usepackage{wrapfig}
\usepackage{subfig}

\usepackage{pstricks,pst-grad,psfrag,pst-node}

\usepackage{caption}
\usepackage{tabularx}
\newcolumntype{x}[1]{>{\raggedright\arraybackslash\hspace{0pt}}p{#1}}
\usepackage{rotating}

\usepackage{cite}

\makeatletter
\newcommand{\logmessage}[1]{\@latex@warning{#1}}
\makeatother
\IfFileExists{../Bibliographie/Users_Guide.txt}{
  }{
  \IfFileExists{../../Bibliographie/Users_Guide.txt}{
    }{
    \IfFileExists{../../../Bibliographie/Users_Guide.txt}{
      }{
        \IfFileExists{../../../../Bibliographie/Users_Guide.txt}{
          }{
            \IfFileExists{../../../../../Bibliographie/Users_Guide.txt}{
              }{
        \logmessage{Directory 'Bibliographie' not found}
          }}}}}

\makeatletter
\newcommand{\ignore}{\logmessage{Text ignored}\@gobble}
\makeatother

\begin{document}
\title{Hybrid tomography for conductivity imaging}

\ifnum\Stiloption=0
	\author{Thomas Widlak${}^1$, Otmar Scherzer${}^{1,2}$}
	\address{${}^1$ Computational Science Center University of Vienna, Nordbergstr. 15, A-1090 Vienna, Austria\\ ${}^2$ RICAM, Austrian Academy of Sciences, Altenberger Str. 69, A-4040 Linz, Austria}
	\ead{thomas.widlak@univie.ac.at}
	\ead{otmar.scherzer@univie.ac.at}
\else
	\author{Thomas Widlak${}^1$, Otmar Scherzer${}^{1,2}$ 
	\\
	\small
	${}^1$ Computational Science Center, University of Vienna, Nordbergstr.~15, A-1090 Vienna, Austria\\ 
	\small
	${}^2$ RICAM, Austrian Academy of Sciences, Altenberger Str.~69, A-4040 Linz, Austria}
	\date{December 13, 2011}

	\pagestyle{empty}
	\maketitle
	\pagestyle{fancy}

	\fancyhf{}
	\renewcommand{\headrulewidth}{0pt}
	\fancyhead[LE,RO]{\thepage}
	\fancyhead[CE]{\uppercase{T. Widlak and O. Scherzer}}
	\fancyhead[CO]{\uppercase{Hybrid tomography for conductivity imaging}}
\fi

\begin{abstract}
Hybrid imaging techniques utilize couplings of physical modalities -- they are called hybrid, because,
typically, the excitation and measurement quantities belong to different modalities. Recently there has been an enormous research
interest in this area because these methods promise very high resolution.
In this paper we give a review on hybrid tomography methods for \emph{electrical conductivity} imaging.
The reviewed imaging methods utilize couplings between electric, magnetic and ultrasound modalities.
By this it is possible to perform high-resolution electrical impedance imaging and to overcome
the low-resolution problem of electric impedance tomography.
\end{abstract}

%%%%%%%%%%%%%%%%%%%%%%%%%%%%%%%%%%%%%%%
%%%%%%%%%%%%%%%%%%%%%%%%%%%%%%%%%%%%%%%
\section{Introduction}

%%%%%%%%%%%%%%%%%%%%%%%%%%%%%%%%%%%%%%%
%%%%%%%%%%%%%%%%%%%%%%%%%%%%%%%%%%%%%%%

\begin{figure}[h]
\subfloat[Low frequency contrast of $\sigma$][Low frequency contrast of $\sigma$ \cite{HaeStaTsaTunMah03,SurStuBarSwa88,KesKesSma06}]
{
\label{fig:Niederfrequenzkontrast}
%\psfrag{Liver}{Liver}
\includegraphics[width=0.45\textwidth,height=0.17\textheight]{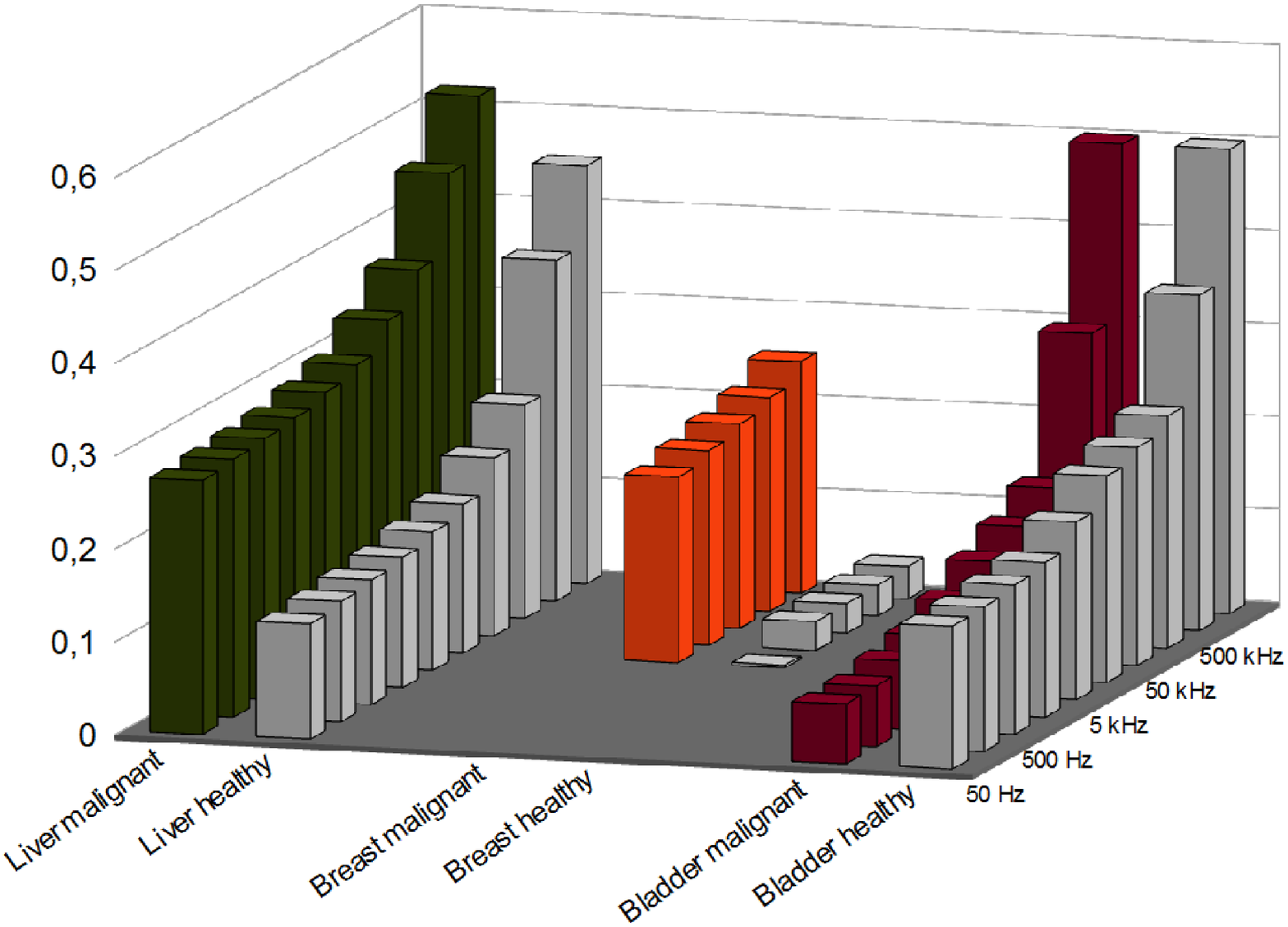}
}
\subfloat[Radio-frequency frequency contrast of $\sigma$][Radio-frequency contrast of $\sigma$\cite{JoiZhaLiJir94}]
{
\label{fig:Hochfrequenzkontrast}
\includegraphics[width=0.45\textwidth,height=0.17\textheight
]{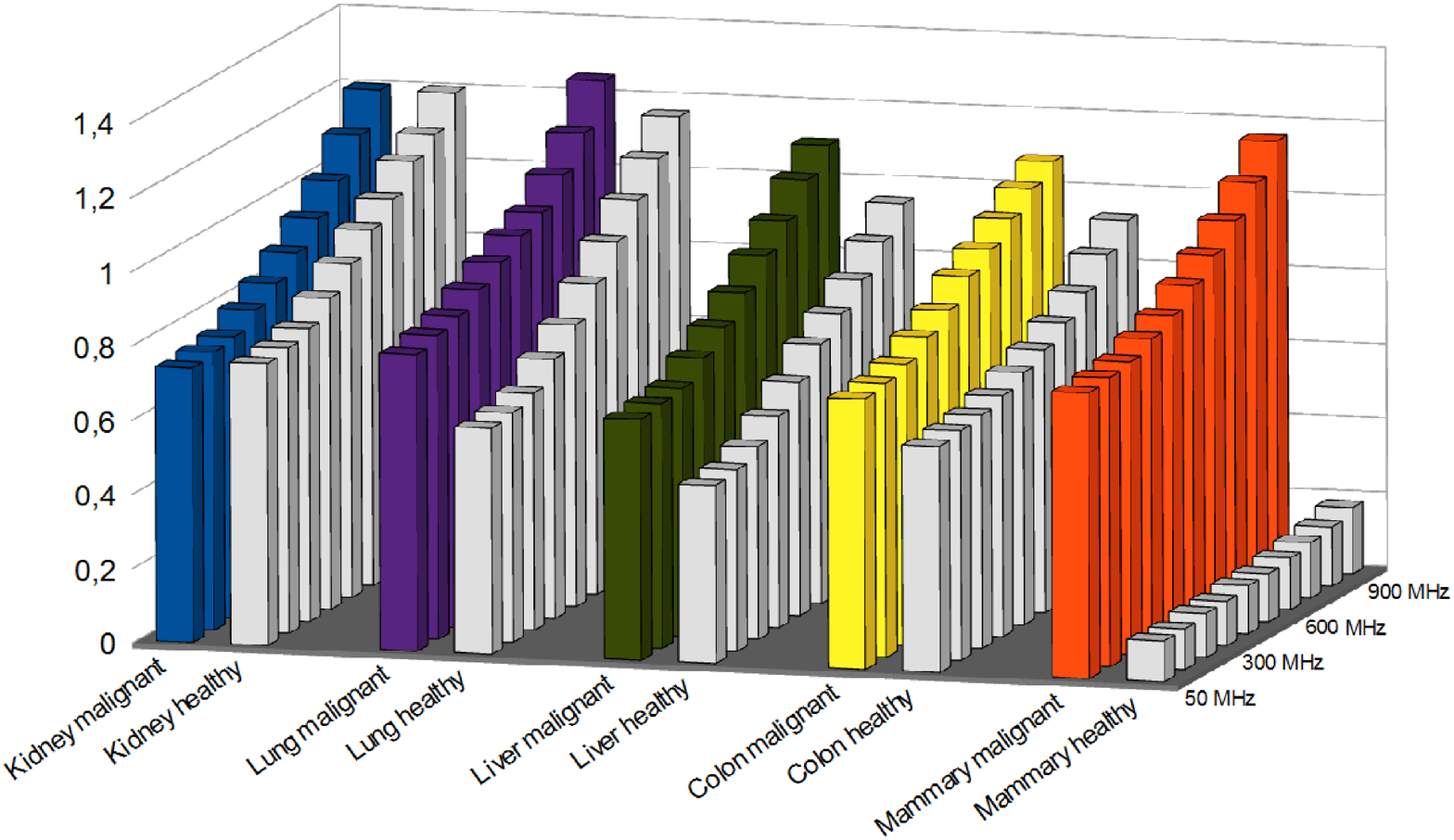}
}
\caption{Contrast of conductivity in biological tissue}
\end{figure}

The spatially varying \emph{electrical conductivity}, denoted by  $\sigma = \sigma(\v x)$ in the following, provides important functional information for diagnostic imaging -- it is an appropriate parameter for distinguishing between malignant and healthy tissue \cite{Schw63,GabLauGab96,HaeStaTsaTunMah03,SurStuBarSwa88,KesKesSma06,MikPavHar06,LauIvoReuRubSol10,Jos98,JoiZhaLiJir94,ZouGuo03}. Although almost never stated explicitly, but very relevant for this work, the conductivity is frequency dependent, that is, $\sigma = \sigma(\v x,\omega)$. The dependence on $\omega$ is typically neglected, since the imaging devices operate at a fixed frequency. However, one should be aware that different image devices record conductivities in different frequency regimes.
In figures \ref{fig:Niederfrequenzkontrast} and \ref{fig:Hochfrequenzkontrast}, respectively, the specific values for $\sigma$ in healthy and malignant tissues are plotted, and it can be observed that cancerous tissue reveals a higher conductivity, in general.

The standard approach for determining the electrical conductivity is \emph{Electrical Impedance Tomography} (EIT). This approach is based on the \emph{electrostatic} equation:
\begin{equation}
\label{Calderon}
\div{(\sigma\grad{u})}=0\,,
\end{equation}
which describe the relation between the conductivity $\sigma$ and the electrostatic potential $u=u(\v x)$. For imaging, EIT is realized by injecting series of currents at a certain frequency $\omega$ on the surface $\partial \Omega$ of the probe and measuring the according series of voltages there. In mathematical terms the EIT problem accounts for computing $\sigma$ from the Dirichlet-to-Neumann mapping $\Lambda_\sigma: f \mapsto g$, which relates the components of all possible pairs $(f=\left.  u\right |_{\partial\Omega}$, $g=\left. \sigma\, \partial_\v{n} u \right|_{\partial\Omega}).$

Precursors of EIT were for geological prospecting \cite{Hum30,Sli33}. Current applications of EIT are in geology \cite{Aiz10,Pel02,SamCouTabBruRic05}, but the applications in non-destructive testing for industry are of growing importance \cite{Wil10}. There has been considerable effort in adapting EIT for \emph{medical} applications \cite{CheIsaNew99,Bay06}.

In mathematical terms, the EIT problem has been formulated first in 1980 by Calder\'on \cite{Cal80}. Uniqueness and stability have been studied extensively (see e.g. \cite{Uhl09,AdlGabLio11,Bor02}).\footnote{ The first step to prove unique recovery of $\sigma$ from $\Lambda_\sigma$ for smooth conductivities, was the result in \cite{SylUhl87}. Now, there exist results for less regular conductivities (see the references in \cite[14.2.1.7]{AdlGabLio11} and \cite[6.1]{Uhl09}). For planar domains, one even has uniqueness for $L^\infty$-conductivities \cite{AstPai06}.} The recovery of the electrical conductivity from the Dirichlet-to-Neumann map is heavily ill-conditioned, the modulus of continuity is at most logarithmic \cite{Ale88, Man01}. This result is consistent with the abilities of classical EIT to differentiate between conductors and insulators in geological prospecting (see table \ref{tab:leitwerte}). On the other hand, the material differences in human and biological tissues are smaller, and logarithmic stability in this case guarantees only very low resolution.

\begin{table}[h]\centering
\begin{tabular}{lc}
tissue type/material&conductivity $\sigma$ \\
\hline
skin (wet) & \unit{3\cdot 10^{-3}}{ \siemens\per\metre}\\
blood &\unit{7\cdot 10^{-1}}{\siemens\per\metre} \\
fat &\unit{2\cdot 10^{-2}}{\siemens\per\metre} \\
liver &\unit{5\cdot 10^{-2}}{\siemens\per\metre} \\
copper &\unit{6\cdot 10^7}{\siemens\per\metre} \\
granite (dry) & \unit{10^{-8}}{\siemens\per\metre}
\end{tabular}
\caption{Conductivities of different type of tissue or materials, respectively, at \unit{1}\kilo\hertz \cite{GabLauGab96}, 
%\cite[Tab. 7.3]{MalPlo95}, 
\cite[Tab. 9.11]{Emi92},
 \cite{DubPiwSanWee78}}\label{tab:leitwerte} \end{table}

Recently, \emph{coupled Physics imaging methods} (this is another name for hybrid imaging, which is more self explaining), have been developed, which can serve as alternatives for estimating the electrical conductivity. \emph{Passive coupled Physics} electrical impedance imaging utilizes that electrical currents in the interior of a probe manifest themselves in different modality. In contrast, \emph{active} coupled Physics imaging is by local excitement of the tissue sample. Mathematically, in most cases, coupled Physics imaging decouples into two inverse problems, which, in total, are expected to better well conditioned than the EIT problem \cite{Amm08}. Coupled Physics imaging led to a series of new results in inverse problems theory (see \cite{Bal11b,Kuc11}).

Coupled Physics imaging should not be confused with \emph{multi-modal} imaging (actually both are sometimes called \emph{hybrid} imaging), which is based on co-registering or fusing images of different modalities. In contrast to multi-modal imaging, the techniques described here, all involve interaction of physical processes, and therefore they have to use the correct modeling of the physics and chemistry involved.

In this paper we survey mathematical models for coupled Physics electrical impedance imaging. We provide the principal mathematical modeling tools in section \ref{sec:Generalmodellierung}, which are various electromagnetic equations in different frequency ranges, basic equations of acoustics, and basics of magnetic resonance imaging (MRI). Based on this, we derive mathematical models for multiphysics imaging models in section \ref{sec:Einzelmodelle}. Bibliographic references are provided, covering experimental and mathematical research.

%%%%%%%%%%%%%%%%%%%%%%%%%%%%%%%%%%%%%
%%%%%%%%%%%%%%%%%%%%%%%%%%%%%%%%%%%%%
\section{Background}
\label{sec:Generalmodellierung}
In the following we provide basic background information on mathematical equations in \emph{Electromagnetism}, \emph{Acoustics}, and \emph{Magnetic Resonance Imaging} (MRI). This overview is required to survey coupled Physics impedance imaging techniques in section \ref{sec:Einzelmodelle}.

\subsection{Electromagnetic theory}\label{sec:Elektromagnet}
We use as a starting point Maxwell's equations, which couple the physical quantities summarized in table \ref{tab:elektromagnetisch}.
\begin{table}
\centering
\begin{tabular}{cll}
symbol&name&unit\\ \hline
$\rho(\v x)$& electrical charge density&\ampere\second\per\cubic\metre \\
$\v E(\v x,t)$ & electrical field strength	& $\volt\per\metre$ \\
$\v H(\v x,t)$& magnetic field strength&$\ampere\per\metre$\\
$\v D(\v x,t)$ & electrical displacement field	& $\ampere\second\per\metre\squared$ \\
$\v B(\v x,t)$& magnetic  flux density&$\volt\second\per\metre\squared=\tesla$ \\
$\v J(\v x,t)$ & electrical current density 	& $\ampere\per\metre\squared$  \\
$\sigma(\v x,\omega)$&electrical conductivity &$\ampere\per\volt\metre=1\per\ohm\metre=\siemens\per\metre$ \\
$\varepsilon(\v x,\omega)$&electrical permittivity&\ampere\second\per\volt\metre=\farad\per\metre\\
$\mu(\v x,\omega)$&magnetic permeability&\volt\second\per\ampere\metre=\henry\per\metre\\
\end{tabular}
\caption{Electromagnetic quantities}\label{tab:elektromagnetisch}
\end{table}

\subsection*{Maxwell's equations}
are the basic formulas of electromagnetism, and read as follows (see for instance \cite{Jac98}):
\begin{equation}\label{eq:Maxwell}
\begin{aligned}
\rot{ \v{E}} &=  -\frac{\partial\v{B}}{\partial t}\,,& \div{\v{D}} &=\rho\,,\\
\rot{ \v{H} }&=\v{J}   + \frac{\partial\v{D}}{\partial t}\,,&\div{\v{B}}&= 0.
\end{aligned}
\end{equation}
For later reference, we recall two equations relating the fields $\v E$, $\v B$ with the charge density $\rho$ and the current density $\v J$:
\begin{itemize}
\item First, the fields $\v E$ and $\v B$ act on free charges and electrical currents by the \emph{Lorentz force}. The force per unit volume is given by (see \cite[(5.12)]{Jac98}):
\begin{equation}
\label{eq:Lorentz}
\v f(\v x,t)=\rho\v E + \v J\times\v B.
\end{equation}
\item Secondly, the work of the electromagnetic fields per unit volume and unit time is the power density \cite[(6.103)]{Jac98}, which in formulas expresses as
\begin{equation}\label{eq:Energie}
H(\v x,t)=\v J\cdot \v E.
\end{equation}
\end{itemize}

\bigskip
In electromagnetism, in many applications, it is observed that the vector fields in \eqref{eq:Maxwell} have the form
\[\begin{aligned}
\v E(\v x,t) &=\left(E_0^k(\v{x}) \cos(\omega t + \varphi_E^k)\right)_{k=1,2,3}\,,\\
\v B(\v x,t) &=\left(B_0^k(\v{x}) \cos(\omega t + \varphi_B^k)\right)_{k=1,2,3}\,,  \\
\v D(\v x,t) &=\left(D_0^k(\v{x}) \cos(\omega t + \varphi_D^k)\right)_{k=1,2,3}\,, \\
\v H(\v x,t) &=\left(B_0^k(\v{x}) \cos(\omega t + \varphi_H^k)\right)_{k=1,2,3}\,, \\
\v J(\v x,t) &=\left(J_0^k(\v{x}) \cos(\omega t + \varphi_J^k)\right)_{k=1,2,3}\;.
\end{aligned}\]

Often the complex phasors
\[
\begin{aligned}
\v{\tilde{E}}(\v{x}) &=\left (E_0^k(\v{x})\, e^{i\omega \varphi_E^k}\right)_{k=1,2,3}\,,\\
\v{\tilde{B}}(\v{x}) &=\left (B_0^k(\v{x})\, e^{i\omega \varphi_B^k}\right)_{k=1,2,3}\,,\\
\v{\tilde{D}}(\v{x}) &=\left (D_0^k(\v{x})\, e^{i\omega \varphi_D^k}\right)_{k=1,2,3}\,,\\
\v{\tilde{H}}(\v{x}) &=\left (H_0^k(\v{x})\, e^{i\omega \varphi_H^k}\right)_{k=1,2,3}\,,\\
\v{\tilde{J}}(\v{x}) &=\left (J_0^k(\v{x})\, e^{i\omega \varphi_J^k}\right)_{k=1,2,3}\,,
\end{aligned}\]
are used to rewrite \eqref{eq:Maxwell} into a system of equations for complex functions without time-differentiation:
\begin{equation}\label{eq:thMaxwell}
\begin{aligned}
\rot{ \v{\tilde{E}}} &=  -   i \omega  \,\v{\tilde{B}}\,,  & \div{\v{\tilde{D} }}    &= \rho \,, \\
\rot{ \v{\tilde{H}}} &= \v{\tilde{J}}+ i\omega\v{\tilde{D}}\,, & \div{\,\v{\tilde{B} }} &= 0.
\end{aligned}
\end{equation}

The real physical fields, which are time-dependent, can be uniquely recovered from the complex but time-independent phasors via the formula $\v E(\v x,t)=\operatorname{Re}\left( \v{\tilde{E}}(\v{x}) \cdot e^{i\omega t}\right)$. We do not distinguish between $\v E(\v x,t)$ and $\v{\tilde{E}}(\v x)$ in the following.

%It is common to assume that the three-dimensional time-dependent vector fields $\v E = \v E(\v x,t)$, $\v D = \v D(\v x,t)$, $\v B = \v B(\v x,t)$ and $\v H = \v H(\v x,t)$ and $\v J = \v J(\v x,t)$ satisfy functional relations \cite[I.4]{Jac98}:
%\begin{equation}\label{eq:Materialcharakteristika1}
%\v J=\v J(\v E,\v B)\,,\qquad\qquad
%\v D=\v D(\v E,\v B)\,, \qquad\qquad
%\v H=\v H(\v E,\v B)\;.
%\end{equation}

For a variety of materials, including biological tissue, the vector fields $\v E$, $\v D$, $\v B$ and $\v H$ and $\v J$ are connected by simple constitutive equations. In these \emph{linear} media, the following  material relations obtain:
\begin{equation}\label{eq:Materialcharakteristika}
	\v{J}  =\sigma      \v{E}\,.		\qquad\qquad	
	\v{D}  =\varepsilon \v{E}\,,         \qquad\qquad
	\v{B}  =\mu         \v{H}\;,
\end{equation}
through which the material parameters $\sigma = \sigma (\v x,t)$, $\varepsilon = \varepsilon(\v x,t)$ and $\mu=\mu (\v x,t)$ are defined. These are in general matrix-valued functions, depending on space and time.

For biological tissue, the parameters  conductivity and permittivity parameters have been investigated thoroughly \cite{Schw63,GabLauGab96}. These parameters are considered mostly as \emph{scalar}, although rarely also anisotropy is investigated. A salient feature of the electrical parameters in tissue is their \emph{dispersion}: That is, $\sigma=\sigma(\v x,\omega)$ and $\varepsilon=\varepsilon(\v x,\omega)$ are functions of the frequency $\omega$ of the electrical field, respectively \cite{Schw63,GabLauGab96}. However, as already stated above, this feature is mostly neglected, since the experiments usually operate in a fixed frequency regime.

In biological tissue the electrical conductivity $\sigma$ varies from $10^{-4}$ to $\unit{10^2}{\siemens\per\metre}$. The permittivity $\varepsilon$ and the permeability $\mu$ are always scalar multiples of the fundamental constants $\varepsilon_0\approx\unit{ 8,9 \cdot 10^{-12} }\farad\per\metre$, $\mu_0\approx\unit{1,3 \cdot 10^{-6}}{\henry\per\metre}$, respectively (see \cite[Tab. A.3]{Jac98}). Moreover, in biological tissue, the ratio $\varepsilon/\varepsilon_0$ is between 10 and $10^8$ \cite{GabLauGab96}. Permeability is usually regarded as a constant \cite[p. 151]{WilKau81}. In the rest of the article, we assume that $\mu=\mu_0$.

\medskip
For the introduction of \emph{potential functions}, we consider the following two equations of Maxwell's system \eqref{eq:thMaxwell}:
 \begin{equation}
\label{eq:FaradayAmpere}
\begin{aligned}
\div{\v B} & =0\,,\qquad \rot{\v E} & =-i\omega \v B\;.
\end{aligned}
\end{equation}
Note, that we identified $\v E$, $\v B$ and $\v{\tilde{E}}$, $\v{\tilde{B}}$, respectively. The first equation already guarantees that there exist a vector potential $\v A$ satisfying
\begin{equation}
\label{eq:Vektorpotential}
\v B=\rot{ \v A}\;.
\end{equation}
Consequently, by using \eqref{eq:Vektorpotential} in the second equation of \eqref{eq:FaradayAmpere}, it follows there also exists a scalar potential $u$, such that
\begin{equation}
\label{eq:allgEPotential} \v  E=- i\omega \v A - \grad{u}.
\end{equation}
Through \eqref{eq:allgEPotential}, the scalar potential $u$ is determined by the observable field $\v E$ up to an additive constant.
Similarly, the vector potential $\v A$ is not uniquely determined by the observable field $\v B$ in \eqref{eq:Vektorpotential}. In fact, one is free to choose  $\div{\v A}$ (see \cite[Thm. in 1.16]{ArfWeb01}), e.g.
\begin{equation}\label{eq:Coulomb}
\div{\v A}=0.\end{equation}
By choosing a particular constant for $u$ and requiring \eqref{eq:Coulomb}, the two equations in \eqref{eq:FaradayAmpere} are equivalent to \eqref{eq:Vektorpotential} and \eqref{eq:allgEPotential}. --
The requirement \eqref{eq:Coulomb} is called \emph{Coulomb's gauge} \cite[6.3]{Jac98}, and we will use it in the rest of the article. 

\subsection{Approximations for lower frequencies}
\label{sec:Approximationen}
In the following we restrict attention to linear media, such as biological tissue. In mathematical terms this means that \eqref{eq:Materialcharakteristika} are satisfied.

\subsubsection*{Quasistatic approximation:} We consider the third equation in \eqref{eq:thMaxwell}, which is called the \emph{Amp\`ere-Maxwell law}. For linear media, it reads
\[\rot {\left(\frac{1}{\mu_0}\v B\right)}=\sigma\v E+ i\omega\varepsilon\v E.\]

If $\omega$ satisfies
\begin{equation}\label{eq:quasistatisch}\omega\ll \sigma/ \varepsilon,\end{equation}
one can neglect the term $i\omega\varepsilon\v E$
(because it is small compared  with $\sigma\v E$). We replace it with the classical version of Amp\`ere's law,
\begin{equation}\label{eq:Ampere}\rot{\left(\frac{1}{\mu_0}\v B\right)}=\v J.\end{equation}
Using \eqref{eq:Ampere} together with the other three equations in \eqref{eq:thMaxwell}, we get the following \emph{quasistatic approximation} to Maxwell's equations:
\begin{equation}\label{eq:thquasistat}
\begin{aligned}
\rot{ \v{E}} &=  -   i \omega \mu_0 \,\v{H}, &  \div{(\varepsilon\,\v{E})}  &= \rho\,,\\
\rot{ \left(\frac{1}{\mu_0}\v{B}\right)} &= \sigma  \,\v{E}, & \div{\v{B}}  &= 0\,.
\end{aligned}
\end{equation}

The quantities
$\v J :=\sigma\v E$ and $\v B$ appearing in \eqref{eq:thquasistat} are related by the \emph{Biot-Savart} law of Magnetostatics \cite[5.3]{Jac98}:
\begin{equation}\label{eq:BiotSavart}
\v{B}(\v{x})= \frac{\mu_0}{4\pi}\int_{\mathbb{R}^3} \v{J}(\v{y})\times \frac{\v{x}-\v{y}}{|\v{x}-\v{y}|^3} d^3\v{y},\qquad\v x\in\mathbb{R}^3\;.
\end{equation}

This law determines the magnetic field which is generated by a current density $\v J$. It is derived from the third and forth equation in \eqref{eq:thquasistat}, i.e. Amp\`ere's law and  $\div{\v B}=0$ (see for example \cite[5.1]{Leh08}). --
Similarly, one can determine the vector potential corresponding to $\v B$ (with Coulomb gauge) as
\begin{equation}\label{eq:BSVP}
\v{A}(\v x)=\frac{\mu_0}{4\pi}\int_{\mathbb{R}^3} \frac{\v{J}(\v y)}{|\v{x}-\v{y}|} d^3\v{y}.
\end{equation}

Certain dispersion models for $\sigma$ and $\varepsilon$, like \cite{GabLauGab96}, suggest that the quasistatic condition \eqref{eq:quasistatisch} holds for a wide range of tissue types even at frequencies about \unit{1-2}{\mega\hertz}. Note, however, that the available data on $\sigma$ and $\varepsilon$ have been obtained using \emph{in vitro} measurements.

At present, condition \eqref{eq:quasistatisch} is often also assumed because it simplifies the subsequent \emph{inverse problem} of determining $\sigma$.
Whether the displacement current should be taken into the model or not depends on the specific application and the available data on $\sigma$ and $\varepsilon$. For frequencies about \unit{1}{\kilo\hertz}, simulation studies have been conducted using a head resp. spine model \cite{WagZahGroPas04,EftSamPol10}, in order to decide whether $\omega\varepsilon\v E$ can be neglected. To the knowledge of the authors, there are at present no studies for higher frequencies, though further empirical and numerical research is necessary to determine the validity of the quasistatic regime in conductivity imaging.

\subsubsection*{Electrostatic approximation:}
We consider the first equation in \eqref{eq:thMaxwell}, also called \emph{Faraday's law}. In linear media, it reads
\begin{equation}\label{eq:Faraday}\rot{ \v{E}} =  -i\omega\v B. \end{equation}

 It can be shown by a dimensional analysis (see \cite[A.2]{CheIsaNew99}) that if
\begin{equation}\label{eq:elektrostatisch}\sqrt{\omega\mu_0\sigma}\ll R,\end{equation}
where $R$ denotes a reference length of $\v x$, then Faraday's law can be replaced by the \emph{electrostatic} equation $\rot{\v E}=0$.
From the latter equation, in turn, it follows that the electrical field can be represented as a gradient field, i.e.,
\begin{equation}\label{eq:Potential}
\v E=-\grad{u}.
\end{equation}

The electrostatic representation \eqref{eq:Potential} is inserted into the complex time harmonic Amp\`ere-Maxwell equation $\rot{ \v{H}}  = \sigma \v{E}+ i\omega\varepsilon\v{E}$ from \eqref{eq:thMaxwell}. Taking the divergence, one gets
\begin{equation}\label{eq:complexCalderon}
\div{(\kappa \grad{u})}=0.
\end{equation}
Here, the quantity \begin{equation}\label{eq:KomplexKappa}\kappa(\v x,\omega)=\sigma(\v x,\omega) +  i \omega \,\varepsilon(\v x,\omega)\end{equation}
is called \emph{complex conductivity} and is measured in $\siemens\per\metre$ as well.

\bigskip
Often, though, one combines the quasistatic with the electrostatic approximation: If we apply the divergence to $\v J$ in the quasistatic version of Amp\`ere's law \eqref{eq:Ampere},  this yields the real-valued elliptical equation in \eqref{Calderon},
\begin{equation}\label{eq:CalderonMaxwell}
\div{\v J} \stackrel{\eqref{eq:Materialcharakteristika}}{=}- \div{ \sigma \v E} \stackrel{\eqref{eq:Potential}}{=}- \div{(\sigma \grad{u})}=0.
\end{equation}
We use later that in the quasistatic case, the magnetic field generated by the current $\v J=\sigma\v E$ is given by the law of Biot-Savart \eqref{eq:BiotSavart}. Similarly, the vector potential generated by $\v J$ can be calculated by \eqref{eq:BSVP}.

Experiments suggest that for $\omega$ larger than \unit{1}{\mega\hertz}, the electrostatic simplifications are not valid anymore and it is reasonable to use a more accurate model like the quasistatic approximation or the whole set of Maxwell's equations \cite{SonPauDehHar06}.

\subsection{A reduced eddy current model}\label{sec:BioWirbelstrom}
In this section we derive a simplified mathematical model of the full Maxwell equations \eqref{eq:thMaxwell} resp. the quasistatic Maxwell equations \eqref{eq:thquasistat}. Such a model was developed in \cite{GenKozIde94} to describe eddy currents in biological tissues (see also \cite{EngSte11}), with the prime goal for improvement of induced current electrical impedance tomography (ICEIT).

\marginpar{
%% http://tex.stackexchange.com/questions/5583/caption-of-figure-in-marginpar-and-caption-of-wrapfigure-in-margin
%%\begin{pspicture}(0,0)(0.0,2.8)
\psfrag{Omega}{$\Large\Omega$}
\psfrag{G}{$ F$}
\includegraphics[width=0.20\textwidth,angle=0,height=0.16\textwidth]{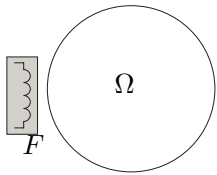}
%%\end{pspicture}
\captionof{figure}{The coil domain $F\supset\operatorname{supp}(\v{J_i})$. which is disjoint from the domain of the conducting specimen, $\Omega$}\label{fig:TraegerAussen}
}

Here, we assume knowledge of a current distribution $\v J_i$ with compact support in a set $F$, where $F$ is a bounded domain outside the conducting material in the domain $\Omega$.
 For instance in ICEIT, $\v J_i$ is the excitation current, and $F$ is the domain of the excitation coil.

We start from the time-harmonic Maxwell equations \eqref{eq:thMaxwell} in the source-free case $\rho=0$. These are simplified below. With an imposed current density $\v{J_i}$, the equations read as follows
\begin{equation}\label{eq:grosseWirbel}
\begin{aligned}
\rot{ \v{E}} &=  -   i \omega \mu_0 \,\v{H}, & \div{(\varepsilon\,\v{E}) } &= 0\,,\\
\rot{ \left(\frac{1}{\mu_0}\v{B}\right)} &= \kappa  \,\v{E}+\v J_i,  &  \div{\v B} &= 0,
\end{aligned}
\end{equation}
where we use the complex conductivity $\kappa$ as in \eqref{eq:KomplexKappa}.

We separate the electrical field $\v E$ into primary and secondary components, and describe those via their potentials (cf. \eqref{eq:allgEPotential}):
\begin{equation}\label{eq:decomp}
\v E = \v{E_p}+ \v{E_s}\qquad \text{ with }\qquad\begin{aligned} \v{E_p} &= -i\omega \v{A_p}\\ \v{E_s}	&= -i\omega \v{A_s}-\grad{u}.\end{aligned}
\end{equation}
Here, $\v A_p$ is the vector potential generated by the imposed current $\v J_i$ (see \eqref{eq:BSVP}). $\v A_p$ is assumed not to depend on the tissue conductivity. The secondary field $\v A_s$ is the vector potential generated by $\sigma\v E$.

For the representation of $\v E_p$ we do not need a scalar potential, whereas for $\v E_s$ we do. Indeed, assume that $\v E_p=-i\omega \v A_p-\grad{u_p}$. Because the permittivity $\varepsilon$ is constant in vacuum, we have $\div(\varepsilon\v E_p)=\varepsilon \div{\v E_p}=0$ (second equation in \eqref{eq:grosseWirbel}).
We use Coulomb's gauge \eqref{eq:Coulomb} for $\v A_p$, yielding
\begin{equation}\label{eq:uCoulomb}
0=\div{\v E_p}=\div{(-i\omega \v A_p-\grad{u_p})}=-i\omega \div{\v A_p}-\div{\grad{u_p}}\stackrel{\eqref{eq:Coulomb}}{=}-\Delta u_p.
\end{equation}
In the source-free case $\rho=0$, the only solutions making physical sense are constant, and without loss of generality. we can take $u_p\equiv 0$ \cite[5.4,6.3]{Jac98}. In particular this means that $\v E_p$ is completely determined by the vector potential $\v A_p$.

Now for biological tissue, it was observed in \cite{GenKozIde94} that
$ \kappa \v{E}	\approx -\kappa ( i\omega \v{A_p} +\grad{u})$. Therefore we can work with
\begin{equation}\label{eq:secundaryA}\v E^{new}:=-i\omega\v A_p-\grad{u}\end{equation} as an approximation for the electrical field. Using the third equation in \eqref{eq:grosseWirbel}, we have $\div{(\rot{\v B})}=\div{\v J}=0$. Consequently, the following condition holds for the scalar potential $u$:
\begin{equation}\label{eq:IC-PoissonKompl}\div{(\kappa\grad{u})}=-i\omega\v A_p\cdot\grad{\kappa}\end{equation}
At the boundary $\partial\Omega$ of the conducting region we have $\kappa \v E^{new}\cdot\v n=0$, as there is no current flux through the boundary. This yields the condition
\begin{equation}\label{eq:PoissonRand}\partial_{\v n} u=-i\omega\v A_p\cdot\end{equation}
at the boundary. Note that in the later applications, the equation \eqref{eq:IC-PoissonKompl} will only be considered on a bounded domain $\Omega$. To ensure uniqueness for the problem \eqref{eq:IC-PoissonKompl}, \eqref{eq:PoissonRand}, one usually requires a scaling condition for $u$, e.g. $\int_{\partial\Omega}u\; dS=0$.

With $\v A_p$ (calculated from the current $\v J_i$) and $u$, we find the electrical field $\v E^{new}$ and a magnetic field $\v B^{new}$ (from \eqref{eq:BSVP}),  which approximately satisfy \eqref{eq:grosseWirbel}. This has been checked in numerical simulations \cite[A.C]{GenKozIde94}, \cite{EngSte11} (see also \cite{DekYinKtiArmPey10}).

For practical simulation and reconstructions, one often restricts oneself to the \emph{quasistatic} version of the eddy current model. That is, one neglects the imaginary part in $\kappa$ and can therefore work with the real electrical conductivity $\sigma$ instead. Then one replaces \label{eq:IC-PoissonKompl} with the model:
\begin{equation}\label{eq:IC-Poisson}
\div{(\sigma\grad{u})}=-i\omega\v A_p\cdot\grad{\sigma}
\end{equation}
As noted in the discussion of the quasistatic condition \eqref{eq:quasistatisch}, which model is to be used depends on the specific application and the available empirical data on the material parameters $\sigma$ and $\varepsilon$.

\subsection{Acoustics}\label{sec:AkustikTheorie}
Equations for the \emph{acoustic} pressure are obtained by linearization from the fundamental equations of fluid dynamics,
which relate the pressure $p=p(\v x,t)$, the density $\rho = \rho(\v x,t)$, and the velocity $\v v = \v v(\v x,t)$ of the fluid, respectively.

Because it is a linearized model the variations of these parameters relative to a ground state $(p_0(\v x),\rho_0(\v x),0)$ \cite[§64]{LanLif87} have to be small to be consistent with reality.

In particular, the standard wave equation
\begin{equation}\label{eq:Wellen}
\square\,p:= \frac{1}{c^2} \partial_{tt} p -\Delta p=0,
\end{equation}
is derived from the linearized conservation principles for impulses, mass, and a relation between pressure and density, which
read as follows:
\begin{align}
\label{eq:linearEuler}\rho_0\partial_t \v v +\grad{p}&=0\,, \\
\label{eq:linearKont}\partial_t \rho + \rho_0 \div{\v v}&=0\,, \\
\label{eq:linDruckDichte}p&=c^2 \rho.
\end{align}
Here, $c=c(\v x)=\sqrt{\partial_\rho p(\rho_0(\v x))}$ denotes the speed of sound in fluid dynamics (see \cite[§ 64]{LanLif87}).
Applying the divergence to \eqref{eq:linearEuler} and calculating the time derivative of \eqref{eq:linearKont}, we
eliminate $\v v $ to get $\Delta p=\partial_{tt}\rho$. Substituting the time derivative of \eqref{eq:linDruckDichte}, we obtain the wave equation \eqref{eq:Wellen}.
\bigskip

In the following, we assume that the fluid is perturbed by electromagnetic fields, causing the Lorentz force \eqref{eq:Lorentz} and Joule heating \eqref{eq:Energie}. We incorporate these effects into the equations for the ultrasound pressure $p$. We first observe that the Lorentz force density $\v f$ enters as a source term into the balance equation for the impulse \eqref{eq:linearEuler}, and we get
\begin{equation}
\label{eq:linearEulerF}
\rho_0\partial_t \v v +\grad{p}-\v f=0.
\end{equation}
Secondly, the \emph{linearized expansion equation}, describes the relation between the energy absorption, described by the power density function $H$, and the pressure
\begin{equation}
\label{eq:linExpansion}
\partial_t p=\Gamma  H + c^2 \partial_t \rho\;.
\end{equation}
Here $\Gamma$ is the dimensionless Gr{\"u}neisen parameter. Very frequently it is assumed constant, but in fact is material dependent, and thus $\Gamma=\Gamma(\v x)$. \eqref{eq:linExpansion} can be derived from thermodynamic relations and the principle of energy conservation \cite[0.1]{Hal10_habil}. We emphasize that \eqref{eq:linExpansion} is a generalization of \eqref{eq:linDruckDichte}.

The wave equation with outer force and energy absorption is now derived from \eqref{eq:linearEulerF}, \eqref{eq:linExpansion} and the mass conservation principle \eqref{eq:linearKont}. We arrive at the following inhomogeneous wave equation
\begin{equation}\label{eq:allgWelle}\square \,p =  -\div{\v f(\v x,t)} + \frac{\Gamma}{c^2}\partial_t H(\v x,t)\end{equation}
We will use this equation with $\v f(\v x,t)=\delta(t)\phi(\v x)$ and $ H(\v x,t)=\delta(t)\psi(\v x)$ in sections \ref{sec:IAT} and \ref{sec:MATMI}.

\begin{remark}\label{sec:RemDuhamel}
By Duhamel's principle (see \cite[p.~81]{Eva98}), the solution of
\begin{equation}\label{eq:InhomogenDuh}
\square \,p = \delta(t) \phi(\v x) + \partial_t \delta(t) \psi(\v x)\,,
\end{equation}
where $\delta$ is the $\delta$-distribution, is the sum of the solutions of the inhomogeneous wave equations
\begin{equation} \label{eq:Duhamel}
\begin{array}{r@{\,}lcl}
\square\, &P_1(\v x,t) &=&0 \\
&P_1(\v x,0)&=&0 \\
\partial_t &P_1(\v x,0)&=&\phi(\v x)\,,
\end{array}
\qquad\qquad
\begin{array}{r@{\,}lcl}
\square\, &P_2(\v x,t) &=&0 \\
&P_2(\v x,0)&=&\psi(\v x) \\
\partial_t &P_2(\v x,0)&=&0\,,
\end{array}
\end{equation}
respectively.
\end{remark}

\section{MRI imaging}
\label{sec:MRI}
\subsection*{General comments and modeling of MRI}
Magnetic Resonance Imaging (MRI) is frequently used as the basis for coupled Physics imaging.

Since the modeling of MRI is not so common in Inverse Problems, it is reviewed here: MRI visualizes the nuclear magnetization $\v M =\v M(\v x,t)$ resulting from selectively induced magnetic fields $\v B = \v B(\v x,t)$. These two quantities are related by the \emph{simplified Bloch equation}\footnote{We restrict our discussion to the influence of $\v B$ on the precessional motion of $\v M$. More adequately, the motion of $\v M$ depends on the material-specific relaxation paramters $T_1$, $T_2$ \cite[4.4]{HaaBroThoVen99}. For the discussion of hybrid techniques, it suffices to study the simplified Bloch equation, and subsequently, the simplified MRI signal in \eqref{eq:MRISig}.}
\begin{equation}\label{eq:Bloch} \partial_t \v M=\gamma\v M\times\v B. \end{equation}
Here $\gamma$ denotes the \emph{gyromagnetic ratio} specific to a proton. For instance, for hydrogen one has $\gamma=\unit{4.6\cdot 10^7}{\reciprocal\tesla\reciprocal\second}$. ($T$ ... Tesla)

For a \emph{stationary} magnetic field in $z$-axis direction, $\v B(\v x,t)=\v B(\v x)=(0,0,B(\v x))$, the solution of the simplified Bloch equation with initial data $\v M(\v x,0) = \v M_0(\v x)$ is given by \cite[2.3.2]{HaaBroThoVen99}
\begin{equation}\label{eq:BlochLsg} \v M(\v x,t)=\begin{pmatrix}  M_0^x \cos(-\gamma B(\v x) t)-M_0^y\sin(-\gamma B(\v x)t)\\
M_0^x \sin(-\gamma B(\v x)t)+M_0^y\cos(-\gamma B(\v x)t)\\ M_0^z\end{pmatrix}.\end{equation}
This identity shows the rotation of the magnetization $\v M$ around the $z$-axis when applying the stationary magnetic field $(0,0,B(\v x))$. The quantity
\begin{equation}\label{eq:Larmor}
\omega(\v x)=-\gamma B(\v x)
\end{equation}
is called the \emph{Larmor frequency}, and is an essential ingredient for MRI.

We explicitly introduce two particular magnetic fields and their Larmor frequency:
\begin{itemize}
\item For the static field \begin{equation}\label{eq:3.2b} \v B_0:=(0,0,B_0),\end{equation}
the Larmor frequency is denoted by
\begin{equation}
\label{eq:static}
\omega_0:=-\gamma B_0.
\end{equation}
For the case of hydrogen, and with $|B_0|=\unit{1.5} {\tesla}$, one has $|\omega_0|=\unit{63.9}{\mega\hertz}$.
\item By a \emph{gradient field} corresponding to the vector $\v G$ we mean the field
\begin{equation}
\label{eq:BG}
\v B^{\v G}(\v x):= (0,0,\v G\cdot\v x).
\end{equation}
If $\v B^{\v G}$ is applied along with $\v B_0$, the Larmor frequency is $\omega(\v x) =  \omega_0- \gamma \v G\cdot\v x$.
\end{itemize}
			
\bigskip

Let now $\v B(\v x,t)=\v B(\v x)$ be a static magnetic field and $\v M(\v x,t)$ the resulting magnetization due to \eqref{eq:Bloch}. The MRI signal is collected from voltage measurements of a detector coil.

It is known that the voltage $V(t)$ received in a detector coil satisfies \[
 V(t)\propto \int \v M(\v x,t)\cdot \v B^{\text{rec}}(\v x) d^3\v x,\]
where $\v B^{\text{rec}}$ is the Biot-Savart-field \eqref{eq:BiotSavart} corresponding to a unit current in the detection coil \cite[7.2]{HaaBroThoVen99}.

Let us assume that $\v B^{\text{rec}}=(r\cos(\theta_B),r\sin(\theta_B),0)$. Then using \eqref{eq:BlochLsg} and trigonometric identities, one finds that the voltage is 
\[
 V(t)= K\int M_0 \sin(\omega(\v x)t+\theta_B-\phi_0)d^3\v x.\]
Here, $M_0(\v x)$ and $\phi_0(\v x)$ are the polar coordinates of the transverse magnetization in the complex representation 
\[M_0 e^{i\phi_0}=M_0^x + i M_0^y.\]
Here, the terms on the right hand side are the $x$- and $y$-components of the magnetization vector $\v M_0=(M_0^x,M_0^y,M_0^z)$.

In a signal analyzing step, one multiplies this voltage with the reference fields $\sin(\omega_0\, t)$ and $\cos(\omega_0\, t)$ respectively. After filtering, ond finds the real and complex part of the following quantity (up to a fixed complex proportionality factor $K$):

\begin{equation} \label{eq:MRISig}
S(t)=\int_{\mathbb{R}^3} M_0(\v x)e^{i\phi_0(\v x)}e^{i\omega(\v x)t} d^3\v x.
\end{equation}
S(t) will be referred to as the MRI signal \cite[23.7]{HaaBroThoVen99}.

%Let now $\v B(\v x,t)=\v B(\v x)$ be a static magnetic field and $\v M(\v x,t)$ the resulting magnetization due to \eqref{eq:Bloch}.
%% Demystifizierung: das ist eine Spannung in einer Detektorspule.
%The actual signal recorded in MRI is \cite[23.7]{HaaBroThoVen99}
%\begin{equation} \label{eq:MRISig}
%S(t)=\int_{\mathbb{R}^3} M_0(\v x)e^{i\phi_0(\v x)}e^{i\omega(\v x)t} d^3\v x.
%\end{equation}
%Here, $M_0(\v x)e^{i\phi_0(\v x)}$ is the transverse magnetization in the complex representation 
%\[M_0 e^{i\phi_0}=M_0^x + i M_0^y,\]
%where the terms on the right hand side are the $x$- and $y$-components of the magnetization vector $\v M_0=(M_0^x,M_0^y,M_0^z)$.

\begin{figure}%\begin{centering}
\begin{centering}
%\begin{pspicture}(0,0)(0.0,5.8)
\psfrag{B0}{$\v B_0$}
\psfrag{B1}{$\v B_1$}
\psfrag{BGz}{$\v B^{\v G_z}$}
\psfrag{BGy}{$\v B^{\v G_y}$}
\psfrag{BGx}{$\v B^{\v G_x}$}
\includegraphics[width=0.5\textwidth]{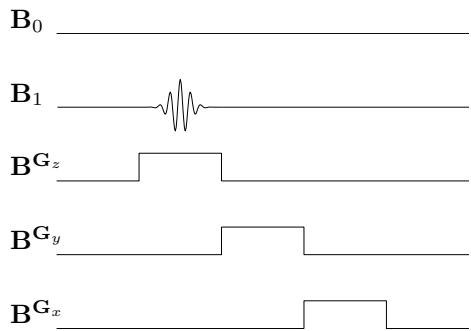}
\caption{Time shape of a pulse sequence with three different gradient fields $\v B^{\v G_x}$, $\v B^{\v G_y}$, $\v B^{\v G_z}$. The signal is measured during application of $\v B^{\v G_x}$.}
\label{fig:MRI}
%\end{pspicture}
\end{centering}
\end{figure}

For MRI imaging, we need only the expression for the signal for the case of a a static field $\v B(\v x)$ as in \eqref{eq:MRISig}, because the signal is only measured while a static field is applied (see figure \ref{fig:MRI}).

\bigskip
The signal $S(t)$ depends both on $\omega(\v x)$ as well as on $\v M$. In actual experiments in MRI, both quantities are varied to collect data.

\subsection*{Experiment:}
In magnetic resonance imaging experiments one applies different fields $\v B$ in order to vary $\omega(\v x)$ and $\v M$, such that sufficient data of $S(t)$ can be collected to image the interior of the specimen. In a general pulse sequence of MRI is a superposition of three magnetic fields, $\v B_0$, $\v B_1$, $\v B^{\v G}$,
\begin{equation}\label{eq:BKomponenten}
\v B(\v x,t)= \v B_0(\v x) + \v B_1(\v x,t) + \v B^{\v G}(\v x,t)\;.
\end{equation}
$\v B_0$ and $\v B^{\v G}$ are as in \eqref{eq:static} and \eqref{eq:BG}, respectively. Moreover, $\v B_1=(B_1(t)\cos(\omega_1 t),B_1(t)\sin(\omega_1 t),0)$ is a radio-frequency field. The time-shapes of the applied fields are depicted in figure \ref{fig:MRI}.

The field $\v B_1$ is responsible for creating a nontrivial $M_0$ as initial data: At every point where the resonance condition
\begin{equation}\label{eq:Resonanz}\omega_1=\omega(\v x)\end{equation}
holds, the magnetization vector is flipped away from the $z$-axis about a certain angle, and inconsequence, a non-zero transverse magnetization $M_0 e^{i\phi_0}$ results.

Gradient fields with different vectors $\v G$ are induced at different time (see figure \ref{fig:MRI}). Applying $\v B^{\v G_z}$ with $\v G_z=(0,0,G_z)$ \emph{during} the radio-frequency field, the resonance condition \eqref{eq:Resonanz} is restricted to a small slice around the plane $z=0$ (and only there $M_0\neq 0$). Every signal \eqref{eq:MRISig} recorded subsequently can be approximated by a 2D integral (cf. \cite[(10.34)]{HaaBroThoVen99}).

One applies $\v B^{\v G_y}$ with $\v G_y=(0,G_y,0)$  \emph{between} the radio-frequency pulse and the measurement for a certain time $t'$. At this time, the Larmor frequency changes to $\omega(\v x)=\omega_0+G_y y$. Afterwards, the phase of the transverse magnetization has changed from $\phi_0$ to $\phi_0+t'G_y y$.

Now after making these preparations, measurements are taken and the signal \eqref{eq:MRISig} is recorded. \emph{During} the measurements, though, a gradient field $\v B^{\v G_z}$ with $\v G_x=(G_x,0,0)$ is applied, which changes the Larmor frequency in \eqref{eq:MRISig} to $\omega(\v x)=\omega_0+G_x x$.

The upshot of applying these three types of gradient fields is that one can modify $\phi_0$ and $\omega(\v x)$ in \eqref{eq:MRISig}, such that one has access to the data
\begin{equation}\label{eq:MRISignalPraeK}
D(tG_x,t'G_y)=\int_{\mathbb{R}^2}M_0 (x,y,0)e^{i\phi_0}e^{i\gamma (tG_x x+t'G_yy)}d(x,y).
\end{equation}
Introducing the 2D vector $\v k=(t G_x ,t'G_y)$, \eqref{eq:MRISignalPraeK} is just the 2D-Fourier transform with respect to $\v k$ of the transverse magnetization $M_0(x,y,0)e^{i\phi_0}$, sampled at the frequency vector $\v k$. By repeating the pulse sequence and varying $(G_x,G_y)$, and thus $\v k$, one collects data in the frequency regime from the Fourier transform of $M_0(x,y,0)e^{i\phi_0}$,
\begin{equation}\label{eq:MRIFourierDaten}
\int_{\mathbb{R}^2}M_0(x,y,0)e^{i\phi_0}e^{i\v k\cdot \v x}d(x,y).
\end{equation}
The Fourier transform allows to recover $M_0(x,y,0)e^{i\phi_0}$.

\begin{remark}
The modeling assumes that the magnetic fields in \eqref{eq:BKomponenten} have an ideal shape, and are not affected by the body tissue, in particular.

In fact, the radio-frequency field $\v B_1$ is affected by electrical properties of the tissue. This effect can be utilized to reconstruct the conductivity $\sigma(\v x,\omega_0)$ using  the Larmor frequency $\omega_0$ \cite{KatDorFinVerNeh07,KatVoiFinVerNeh09, ZhaZhuHe10}. -- This is one of several examples of imaging the electrical conductivity, or the current density at the Larmor frequency, using the measurement setup of MRI (for other approaches see \cite{NegTonCon11,WanMaDeMNacJoy11}).
\end{remark}

\section{Coupled Physics models for conductivity imaging}
\label{sec:Einzelmodelle}

%%%%%%%%%%%%%%%%%%%%%%%%%%%%%%%%%%%%%
%%%%%%%%%%%%%%%%%%%%%%%%%%%%%%%%%%%%%

In this section we review \emph{coupled Physics} techniques for \emph{imaging} the scalar conductivity $\sigma$ from \eqref{eq:Materialcharakteristika}.

\subsection{Magnetic coupling}

Here, we discuss two sorts of methods for conductivity imaging based on magnetic coupling. One set of methods uses coupling of magnetic and electrical phenomena described by Maxwell's equations. The other set of methods uses MRI and exploits the influence of electrical currents on the MRI signal. Both kinds of methods have either contactless excitation or contactless measurement (or both). The latter is a practical advantage in clinical experiments.

\subsubsection{Magnetic resonance-EIT (MREIT) / Current density impedance imaging (CDII):}
\label{sec:MREIT}
In MREIT and CDII, respectively, one perturbs the MRI signals by injecting an electrical current.

\subsubsection*{Experiment:}
The measurement setup consists of an MRI machine and, additionally, electrodes on the surface of the specimen (within the MR machine). The MRI pulse sequences are performed, and an electrical current is injected for a short period, before the induction measurement. The injection perturbs the Larmor frequency and leads to a change in the nuclear magnetization, which alters the MRI signal. In CDII, in addition, the experiment is repeated with the probe rotated relative to the MRI machine.

\subsubsection*{Mathematical modeling:}
The mathematical modeling of these two imaging techniques is very similar to the modeling of MRI in section \ref{sec:MRI}. The difference is that the injected electrical currents manifest themselves in an additional magnetic field, namely, the Biot-Savart field $\v B_{\v J}=(B_{\v J}^x,B_{\v J}^y,B_{\v J}^z)$ from \eqref{eq:BiotSavart}. This field has to be taken into account in \eqref{eq:BKomponenten}, in the mathematical model of MRI. Thus the total magnetic field in MREIT/CDII is given by
\[ \v B(\v x,t)=\v B_0 + \v B_1(\v x,t) + \v B^{\v G}(\v x,t) + \boxed{\v B_{\v J}(\v x,t)}. \]

The current $\v J$ is injected \emph{before} the induction measurement. As $\v J$ is a direct current or has very low frequency, the Biot-Savart field $\v B_{\v J}$ acts like an additional gradient field $\v B^{\v G_y}$. As discussed in section \ref{sec:MRI}, this results in a change of the Larmor frequency, $\Delta\omega=-\gamma |\v B_{\v J}|$, at the time $T_{\v J}$ when the current $\v J$ is applied. The transverse magnetization therefore accumulates an additional phase difference $\Delta\phi=\gamma|\v B_{\v J}| T_{\v J}$, which in this situation is approximately $\gamma B^z_{\v J} T_{\v J}$ (see \cite[27.3.6]{HaaBroThoVen99}).

Compared with the signals \eqref{eq:MRIFourierDaten} recorded in MRI, the signals in MREIT/CDII reveal a phase shift. Because in the experiments two currents of opposite sign are applied \cite{ScoJoyArmHen92,SeoWoo11,WooSeo08}, one collects the following data:
\begin{equation}\label{eq:MREITSignalK}
 S_\pm(k_x,k_y)=\int_{\mathbb{R}^2}
 M_0(x,y,0)\,e^{i\varphi_0}\;
 e^{\pm i\gamma B^z_{\v J}(x,y,0)\,T_{\v J} }\;
 e^{i\v k\cdot \v x}d(x,y),
\end{equation}
Here, $\pm$ indicates two currents of opposite signs. Using the inverse Fourier transforms of \eqref{eq:MREITSignalK}, and denoting
\begin{equation}\label{eq:MRI-Phasengleichung}
m_\pm=M(x,y,z_0)\;
 e^{\pm i\gamma B^z_{\v J}(x,y,z_0)\,T_{\v J} },
 \end{equation}
one finds that:
\begin{equation}\label{eq:BExtrakt}
B^z_{\v J}(x,y,z_0)=\frac{1}{2\gamma T_{\v J}} \operatorname{Im}\left(\log\left(\frac{m_+(x,y,z_0)}{m_-(x,y,z_0)}\right)\right)\,,
\end{equation}
which is the component of $\v B_{\v J}$ in direction of the \emph{main} magnetic field $\v B_0$.

Other components of $\v B_{\v J}$ can be obtained by rotating the sample relative to the MRI machine. Then one determines the current density by Ampère's law \eqref{eq:Ampere} applied to $\v B=\v B_{\v J}$, that is, $\v J=\frac{1}{\mu_0}\rot{\v B_{\v J}}$. 

\subsubsection*{Imaging:}
$\v B^{\v J}$ and $\v J$ depend non-linearly on the conductivity $\sigma$. %The reconstruction of $\sigma$ is the hybrid imaging problem.
Depending on the acquired MRI data it is common to differ between MREIT, that uses values of $B^z$, and CDII, where knowledge of the current density $\v J$ is assumed. CDII, in order to obtain the three dimensional vector $\v B^{\v J}$, requires rotating the probe relative to the MRI-machine three times.

\begin{prob}[CDII]\label{prob:CDII}
For a given domain $\Omega$, representing the domain of the probe to be imaged, let boundary data $f$, $g$ be given. Moreover assume that on the whole of $\Omega$, $\frac{1}{\mu_0}\rot{\v B_{\v J}}=\v J=-\sigma\grad{u}$, or $|\v J|$, respectively, are given, where
$u$ and $\sigma$ are related by
\begin{equation}
\label{u_equation}
\begin{aligned}
\div{(\sigma \grad{u})} &=0 \text{ in }\Omega\subset\mathbb{R}^3\\
u&=f \text{ on }\partial\Omega \\
\sigma\,\partial_\v n u&=g \text{ on }\partial\Omega.
\end{aligned}
\end{equation}
The imaging problem consists in determining $\sigma$.
\end{prob}

\begin{prob}[MREIT]\label{prob:MREIT1}
Let boundary data $f$, $g$ be given. Let $\sigma$ and $u$ be related by (\ref{u_equation}).
The imaging problem consists in determining $\sigma$ from $\v B_{\v J}$, which is related to $\sigma$ and $u$ via the
Biot-Savart relation \eqref{eq:BiotSavart},
\[
\v{B}_{\v J}(\v{x})= -\frac{\mu_0}{4\pi}\int_{\mathbb{R}^3} \sigma(\v y)\grad{u}(\v y)\times \frac{\v{x}-\v{y}}{|\v{x}-\v{y}|^3} d^3\v{y}\;.
\]
\end{prob}
MREIT/CDII have been suggested in the late 1980ies \cite{ScoJoyArmHen91}. Since then, efficient algorithms have been developed for solving the two problems. There also exist a variety of experimental biomedical studies \cite{WooSeo08,SeoWoo11}. In practice, the experiments are repeated at least two times with different boundary data, which provides more independent information on $\sigma$.

\subsubsection*{Mathematics of MREIT and CDII:}
\begin{description}
\item{For CDII, many uniqueness and stability results have been obtained:}
\begin{itemize}
\item Let $|\v{J_1}|$, $|\v{J_2}|$ and corresponding Neumann data $\sigma\,\partial_\v n u=g_i,i=1,2$ be given, where the currents satisfy $\v{J_1}\nparallel\v{J_2}$ everywhere in $\Omega$. Then, $\sigma$ can be recovered uniquely \cite{KwoWooYooSeo02,KimKwoSeoYoo02,KimKwoSeoWoo03}.
\item For two given vector-valued current densities, there exists an analytical reconstruction formula, which determines $\sigma$ up to a multiplicative constant \cite{Lee04,HasMaNacJoy08}.
\item With one current density vector $\v J$ and Dirichlet data $u|_{\partial\Omega}$ given, unique recovery of $\sigma$ for the two-dimensional version of problem \ref{prob:CDII} is guaranteed \cite{KwoLeeYoo02}.
\item Recently, it was observed that the absolute value of one current, $|\v J|,$ and Dirichlet data \cite{NacTamTim09,NacTamTim10} are sufficient for recovering $\sigma$. These uniqueness results for CDII have been proven by interpreting level sets of $u$ as \emph{minimal surfaces} or by exploiting an energy-minizing property of $u$ \cite{NacTamTim07,NacTamTim09,NacTamTim10,NacTamTim11}. They require $u$ to be free of singularities. Stability results of the $|\v J|$ based reconstruction have been given by \cite{NasTam11,NacTamTim07}.
\end{itemize}
\item{For MREIT:}
\begin{itemize}
\item
Uniqueness and stability results for MREIT are given in special cases by \cite{LiuSeoSinWoo07,LiuZhoHe09}.
\item Various numerical algorithms have been developed, which reconstruct $\sigma$ using $B^z_{\v J_i}$ and Neumann data $g_i$, $i\geq 2$. For an overview, see \cite[Sec. 3]{SeoWoo11}.
\end{itemize}\end{description}

\subsubsection{Magnetic detection impedance tomography (MDEIT):}
\label{sec:MDEIT}

MDEIT uses magnetic boundary detection to recover currents and conductivity inside the tissue.

\subsubsection*{Experiment:}
The equipment are electrodes to inject currents and several coils to record the magnetic field. A current is injected into the probe and the strength of the magnetic field is measured by detector coils at different positions of the boundary. This technique is discussed in \cite{IreTozBarBar04,KreKuhPot02} and for practical applications see \cite{LusReiSte09}.

\subsubsection*{Mathematical modeling:}
The mathematical model is the electrostatic model, $\div{(\sigma{\grad{u}})}=0$ from \eqref{eq:CalderonMaxwell}, which relates the electrical potential $u$ and the conductivity $\sigma$. Neumann boundary conditions are given by the injected currents. The Biot-Savart law \eqref{eq:BiotSavart} provides the relation between $\v J=-\sigma\grad{u}$ and the magnetic field $\v B$ measured by the detector coils.

\subsubsection*{Imaging:}
The imaging  problem can be decomposed into two subproblems. In the first part, one reconstructs the current density $\v J$ inside $\Omega$ from measurements of $\v B$ at the boundary $\partial \Omega$. In the second part, one faces exactly the same problem as in CDII, namely to reconstruct $\sigma$ from interior data of $\v J$ (problem \ref{prob:MREIT1}).

The inverse problem in the first step is a linear one:  to recover $\v J$ given the Biot-Savart law. We can restrict ourselves to current densities satisfying $\div{\v J}=0$, see  \eqref{eq:CalderonMaxwell}.

\begin{prob}[Inverse problem MDEIT, first step]\label{prob:MDEIT} 
Determine $\v{J}|_\Omega$ from $\v B|_{\partial\Omega}$ using the Biot-Savart relation \begin{equation}
\label{eq:Biosavart_j}
\v{B}(\v{x})= \frac{\mu_0}{4\pi}\int_{\mathbb{R}^3} \v{J}(\v{y})\times \frac{\v{x}-\v{y}}{|\v{x}-\v{y}|^3} d^3\v{y},\qquad\v x\in\mathbb{R}^3,
\end{equation} taking into account that $\div{\v J}=0$.
\end{prob}

The second problem is analogous to CDII (problem \ref{prob:MREIT1}):
\begin{prob}[Inverse problem MDEIT, second step]\label{prob:MDEIT2}
Let be given data $f$ and $g$ on $\partial \Omega$. Determine $\sigma$ from $\v J=-\sigma\grad{u}$, which are related by (\ref{u_equation}).
\end{prob}

\subsubsection*{Mathematical analysis:}
The mathematical investigations relating to the linear problem \ref{prob:MDEIT} concern uniqueness issues and development of numerical algorithms. The kernel of the Biot-Savart operator, has been investigated in \cite{KreKuhPot02,HauKuhPot05}. Uniqueness in the reconstruction problem has been shown for particular cases, such as directed currents in layered media \cite{KreKuhPot02,HauKuhPot05,PotWan09}. Different regularization methods have been applied to solve problem \ref{prob:MDEIT}, and tested using real and simulated data \cite{IreTozBarBar04,HauPotWan08}.

\subsubsection{EIT with induced currents (ICEIT)}
\label{sec:ICEIT}
ICEIT excites currents by magnetic induction and generates a voltage, which is measurable at the boundary of the probe.

\subsubsection*{Experiment:}
For the experiments, one uses coils and electrodes. A time harmonic current is produced in the excitation coil and eddy currents are induced in the tissue. The voltages are detected at the boundary of the probe \cite{GenKozIde94,ZloRosAbb03}.

\subsubsection*{Modeling:}
We recall the setting of the reduced eddy current model in section \ref{sec:BioWirbelstrom}, referring to figure \ref{fig:TraegerAussen}. The excitation coil lies in a domain $F$ where the excitation current $\v J_i$ is supported. The conducting tissue is in the domain $\Omega$, disjoint from $F$.

In \eqref{eq:secundaryA}, we have introduced a reduced version of the electrical field as
\[\v E=-i\omega\v A_p-\grad{u},\]
where $\v A_p$ is the primary vector potential generated by the excitation current $\v J_i$ (see \eqref{eq:BSVP}). The scalar potential is governed by \eqref{eq:IC-Poisson} with boundary condition $\v E\cdot \v n=0$ on $\partial\Omega$. To ensure a unique solution, we require that $\int_{\partial\Omega}u dS=0$. This gives the description of the eddy currents inside the tissue.

The measurement data consist of (finitely many) line integrals $\int_\v{x_1}^\v{x_2}\v E\cdot d\v l$. Because the determinatino of $\v A_p$ via \eqref{eq:BSVP} does not depend on inhomogeneities of $\sigma$, one obtains line integrals of $\grad{u}$, which are potential differences. As in classical EIT, these correspond approximately to knowledge of Dirichlet data $u|_{\partial\Omega}$, if one chooses a voltage reference. For this, we assume $\int_{\partial\Omega}u\; dS=0$.

\subsubsection*{Imaging:}
The imaging problem in ICEIT is an inverse boundary value problem:
\begin{prob}[Inverse problem ICEIT]\label{prob:ICEIT}
Let $(\sigma,u)$ satisfy
\begin{equation}\begin{aligned}\label{eq:ICEITPoisson}
\div{(\sigma \grad{u})}&= -i\omega\v{A_p}\cdot\grad{\sigma} \text{ in }\Omega \\
\partial_{\v n} u&=-i\omega\v{ A_p}\cdot \v n \text{ on }\partial\Omega\\
\int_{\partial\Omega}u \;dS&=0.
\end{aligned}\end{equation}
Given are boundary data $u|_{\partial\Omega}$ and the primary potential $\v{A_p}|_\Omega$ in $\Omega$, which is related to the excitation current $\v{J_i}$ on $\mathbb{R}^2$ by
 \begin{equation}\label{eq:primaryA}
\v{A_p}(\v x)=\frac{\mu_0}{4\pi}\int_{\mathbb{R}^2} \frac{\v{J_i}(\v y)}{|\v{x}-\v{y}|} d^3\v{y}.
\end{equation}
Determine $\sigma$.
\end{prob}

\subsubsection*{Mathematics and Numerics:}
ICEIT has been investigated numerically: Different regularization procedures have been applied, e.g., in \cite{GenKozIde94,ZloRosAbb03}.

\subsubsection{Magnetic induction tomography (MIT):}
\label{sec:MIT}
In MIT, eddy currents are induced, and the resulting magnetic fields are measured at the boundary.
\subsubsection*{Experiment:}
The measurement equipment consists of coils placed at the boundary of the conducting tissue. Electrical currents in the coils induces eddy currents in the tissue. These currents generate a secondary magnetic field, which is measured by detection coils \cite{SteLeo07,GurScha09}, outside of the probe.

\subsubsection*{Modelling:}
The mathematical model is based on the reduced eddy current model (\ref{eq:grosseWirbel}) in section \ref{sec:BioWirbelstrom}, as in the previous section.

By \eqref{eq:secundaryA}, the electrical field is represented as $\v E=-i\omega \v A_p-\grad{u}$, where $\v A_p$ is the primary vector potential, generated by the excitation current $\v J_i$ (see \eqref{eq:BSVP}). The scalar potential $u$ is determined by \eqref{eq:IC-Poisson}.

The signal recorded in a detection coil is the electrical voltage induced by $\v E|_\Omega$. Special types of detectors can be constructed, which are not influenced by the primary excitation field $\v E_p=-i\omega\v A_p$, but only by the reaction field \cite{GurScha09}. As in \eqref{eq:decomp}, the electrical reaction field is $\v E_s=-i\omega\v A_s-\grad{u}$, where $\v A_s$ is the vector potential generated by the current $\sigma\v E$ (see also \cite[(25)]{EngSte11}).

The measured voltage in a coil with surface $C$ is then
\[v=\oint_{\partial C} \v{E_s}\cdot d\v l=\int_C \v n\cdot(\rot{\v{E_s})}=-i\omega\int_C \v n\cdot\v{B_s}\;.\]
In the last equation, we have applied Faraday's law \eqref{eq:Faraday} to $\v E=\v E_s$ and $\v B=\v B_s$. $\v{B_s}$ is the magnetic field generated by the current current
\[\sigma\v{E}|_\Omega=-i\sigma\omega \v{A_p}-\sigma\grad{u}\;,\]
see \eqref{eq:BiotSavart}.

\subsubsection*{Imaging:}
The reconstruction problem of MIT is as follows:
\begin{prob}[Inverse problem MIT]\label{prob:MIT}
Let $\v{A_p}$ be the vector potential generated by the excitation current $\v{J_i}$, given by \eqref{eq:primaryA}. Let $u$ satisfy the Poisson equation \eqref{eq:ICEITPoisson} with parameter $\sigma$.

The given data are different voltage measurements
\[V_k=\int_{C_k} \v n\cdot\v{B_s}\]
acquired with detector coils on the surface $C_k$. Here, $\v{B_s}$ is function of $\sigma$, $u$ and $\v{A_p}$ as \begin{equation}
\label{eq:MsecundaryB}
\v{B}_s(\v x)=-\frac{\mu_0}{4\pi}\int _\Omega \sigma\left(i\omega\v{A_p}(\v y)+\grad{u}(\v y)\right)\times\frac{\v x-\v y}{|\v x-\v y|^3} \,d^3\v{y}, \qquad \v x\in\mathbb{R}^3.
\end{equation}
The imaging problem consists in determining $\sigma$.
\end{prob}

\bigskip
Eddy current imaging techniques play an important role for non-destructive evaluation in industry \cite{Tak97}. Actually, magnetic induction tomography has been developed for industrial non destructive evaluation and geological surveying \cite{MorKun62,PeyYuLyoAlZFer96,BinLyoPeyPri01}. For these application the reduced model of magnetic induction is not valid. However, also biomedical applications of MIT have been proposed later and the reduced model is now in general use. But MIT does not provide high resolution \cite{GurScha09}. The primary importance of biomedical MIT is for contactless monitoring of physiological processes such as respiration \cite{SteLeo07}. However, it was observed recently that coupling MIT with ICEIT (sect. \ref{sec:ICEIT} and classical EIT leads to better resolution of the corresponding linearized inverse problem \cite{GurMamAdlScha11}.

\subsubsection{MREIT with magnetic induction (ICMREIT) / Current density impedance imaging with magnetic induction (CDII-MI):}
\label{sec:ICMREIT}
In ICMREIT and CDII-MI, one perturbs the MRI signals by an electrical current, which is produced by magnetic induction.

\subsubsection*{Experiment:}
The measurement apparatus is an MRI machine with one or more coils for inducing eddy currents in the probe. MRI pulse sequences are performed. Additionally, eddy currents are produced before the induction measurement, and the perturbation of the MRI signal is recorded.

 The technique has been first proposed in \cite{OzpIde05}.

\subsubsection*{Modeling:}
The modeling of these techniques is similar to MREIT/CDII discussed in section \ref{sec:MREIT}, except that we use the reduced eddy current model \eqref{eq:IC-Poisson} from section \ref{sec:BioWirbelstrom} to model the electric field.

By \eqref{eq:secundaryA}, the eddy current inside the tissue is
$\v {J}= -\sigma (i\omega \v{A_p}+\grad{u}),$
with $\v A_p$ determined as vector potential generated by the impressed current density $\v J_i$. This (low frequency) eddy current generates a magnetic field component $\v B_{\v J}$ as in \eqref{eq:MsecundaryB}.

As discussed in the section \ref{sec:MREIT} on MREIT/CDII, the signal measured has information of one component of the magnetic field, as in \eqref{eq:BExtrakt}; this information is used in ICMREIT. By rotating the probe, one can obtain the full vector field $\v B_{\v J}$, which is used in CDII-MI.

\begin{prob}[Inverse problem CDII-MI]\label{prob:CDIIMI}
Let $\v{A_p}$ be the vector potential generated by the excitation current $\v{J_i}$, given by \eqref{eq:primaryA}. Let $u$ satisfy the Poisson equation \eqref{eq:ICEITPoisson} with parameter $\sigma$.

Given are data $\v{J}=-\sigma (i\omega \v{A_p}+\grad{u})$.
Determine $\sigma$.
\end{prob}

\begin{prob}[Inverse problem ICMREIT]\label{prob:ICMREIT}
Let $\v{A_p}$ be the vector potential generated by the excitation current $\v{J_i}$, given by \eqref{eq:primaryA}. Let $u$ satisfy the Poisson equation \eqref{eq:ICEITPoisson} with parameter $\sigma$.

Given are data $\v{B}_{\v J}^z$ with $\v B_{\v J}$, as in \eqref{eq:MsecundaryB}.
Determine $\sigma$.
\end{prob}

\subsubsection*{Mathematics and Numerics:}
Algorithms for MREIT/CDII have been adapted for the numerical solution of these problems. Thereby using measurements involving $B^z_{\v J}$ \cite{OzpIde05,GeeCreDup10}, or measurements involving $\v J$ \cite{LiuZhoHe09}.

\subsection{Coupling involving acoustic modalities}
\label{sec:AkustikModelle}
Up to now, we have discussed how the conductivity distribution behaves under influence of electric and magnetic fields. In the last ten years, there have been several proposals to utilize couplings between electrical and acoustical properties. Ultrasound propagation translates well from outer measurements to inner quantities of the probe. It has long been used well for other hybrid techniques such as photo-acoustic imaging. Therefore high resolution can be expected. We explain the proposals in the following, and draw especially on background in section \ref{sec:AkustikTheorie}.

\subsubsection{Impedance-acoustic tomography (IAT):}
\label{sec:IAT}
Impedance-acoustic tomography exploits Joule heating due to electric currents to produce an ultrasound wave.

\subsubsection*{Experiments:}
The measurement equipment are electrodes and ultrasound transducers. The electrodes are attached to the surface of the tissue, and an electric current is injected into the tissue. Joule heating results, and the tissue emits an ultrasound wave due to thermal expansion. Ultrasound transducers record this pressure wave at the surface of the probe or in a contact medium, e.g. water. The method has been proposed in \cite{GebSch08}.

\subsubsection*{Modeling:}
We use the quantitative description of the ultrasound wave incorporating electric effects from section \ref{sec:AkustikTheorie}. Since there is no additional outer force $\v f$, the relevant wave equation is

\[\frac{1}{c^2} \partial_{tt} p(\v{x},t)- \Delta p(\v{x},t)=\frac{\Gamma}{c^2}\partial_t H,\]
The model has to be supplied with the initial data $p(\v x,t)=0$ and $\frac{\partial }{\partial t} p(\v x,t)=0$. The equation can be considered in $\mathbb{R}^n\times (0,\infty)$, if one notes that $H$ has its support in $\Omega$.

We assume that $H(\v x,t)=\bar H(\v x) g(t)$, and that $g(t)$ can be assumed as delta pulse. This is because the time of the electrical pulse multiplied with the sound speed in the tissue is negligible. We therefore obtain an inhomogeneous wave equation as outlined in remark \ref{sec:RemDuhamel} in section \ref{sec:AkustikTheorie}:
\begin{equation}
\begin{aligned}\label{eq:IAT}
\frac{1}{c^2} \partial_{tt}p -\Delta p\, &= 0\,, && (\v{x},t) \in&\mathbb{R}^n&\times(0,\infty)\,, \\
p&=\frac{\beta}{C_p}\sigma|\v {\overline E}|^2\,,  &&  (\v{x},t) \in&\mathbb{R}^n&\times\{0\}\,,\\
\partial_{t} p&=0\,, && (\v{x},t) \in&\mathbb{R}^n&\times\{0\}\,,
\end{aligned}
\end{equation}
where $\sigma$ vanishes outside the domain $\Omega$.
%Then the initial pressure has finite support.

The power density $H$ can be related to the conductivity with Joule's law \eqref{eq:Energie}, that is,
\begin{equation}
 \label{eq:LeistungsDichte} H=\v{J}\cdot\v{E}=\sigma|\v E|^2,\end{equation}
with unit $\watt\per\metre\cubed$, quantifying the energy converted per unit volume and unit time. The power density $H$ is the intermediate link between the electrical excitation and the acoustical detection, and it appears in the wave equation as a source term.

\medskip

\subsubsection*{Imaging:}
Recovery of the initial pressure from boundary data is a long-established technique. Reconstruction is possible with different measurement geometries, and it has been mainly developed for photo-acoustic tomography, a hybrid technique combining laser excitation and ultrasound detection \cite{KucKun11,LiWan09}.

If we assume that the quantities $\beta$ and $C_p$ are  constants, then this gives knowledge of the power density $\sigma|\v{\overline{E}}|^2$.

The second step is the reconstruction of conductivity from the interior data $\sigma|\v E|^2=\v J\cdot \v E$ throughout $\Omega$. Obviously, we have to indicate how the electrical field entering this problem as initial source state is to be described. Here it makes a difference which kind of frequency is employed.

In the electrostatic case of impedance-acoustic tomography \cite{GebSch08}, we can use the approximation \eqref{Calderon} known from classical impedance tomography, including Dirichlet data. Then we end up with the following inverse problem with interior data:

\begin{prob}[IAT, conductivity reconstruction]\label{prob:IAT}
Let
\[\begin{aligned}\div{ (\sigma \grad{u})}&=0 \text{ in } \Omega\,,\\
u|_{\partial\Omega}&= f \text{ on } \partial\Omega\,,\\
\sigma\,\partial_\v nu&= g \text{ on } \partial\Omega\;.\end{aligned}\]
Given are $f$, $g$, as well as interior data $H=\sigma|\v \grad{u}|^2$. Determine the conductivity $\sigma$ from these.
\end{prob}

\subsubsection*{Mathematical analysis:}
Unique recovery of $\sigma$ was shown in \cite{CapFehGouKav09} for the two-dimensional version of problem \ref{prob:IAT}, using $H_{11}$, $H_{12}$, $H_{22}$ for $H_{ij}=\sigma\grad{u_1}\cdot\grad{u_2}$. (The term $H_{12}$ can be obtained using $\sigma|\grad{u_1}|^2$, $\sigma|\grad{u_1}|^2$, $\sigma|\grad{(u_1+u_2)}|^2$ by polarization.) This result was generalized in \cite{BalBonMonTri11} to three dimensions, using multiple measurement data to guarantee unique global reconstruction. Reconstruction of $\sigma$ using one power density $H=\sigma|\grad{u}|^2$, based on the nonlinear PDE $\div{ (H\frac{ \grad{u}}{|\grad{u}|^2})}=0$ was investigated in \cite{Bal11a}. Numerical reconstruction procedures have been applied to simulated data in \cite{CapFehGouKav09,GebSch08, KucKun11b}. 
The stability result from \cite{KucSte11} applies to reconstructions of $\sigma$ from two or more power densities.

\subsubsection{Quantitative thermo-acoustic tomography (qTAT):}
\label{sec:qTAT}
In thermo-acoustic tomography, one exploits Joule heating due to microwave excitation to produce an ultrasound wave.

\subsubsection*{Experiment:}
The measurement equipment consists of a microwave generator and ultrasound transducers. Microwaves pulses are generated with a frequency in the {\giga\hertz} range. They propagate within the tissue sample and energy absorption causes Joule heating. The tissue reacts with a short expansion wave which is recorded outside the tissue, either on its surface or in a contact medium such as water. Measurement setups are described, e.g., in \cite{LiWan09}.

\subsubsection*{Modeling:}
The general principles of the technique are the same as for \emph{impedance-acoustic tomography}, described in section \ref{sec:IAT}. We use again the inhomogeneous wave equation \eqref{eq:Duhamel} with $\v f=0$. Assuming that the power density is of the form $H(\v x,t)=\bar H(\v x)g(t)$ with $g(t)=\delta(t)$ a delta pulse. Therefore, we arrive at the same wave equation \eqref{eq:IAT} with initial pressure $\bar H(\v x)=\v J(\v x)\cdot\v E(\v x)=\sigma(\v x)|\v E(\v x)|^2$.

A high-frequency description is necessary for modelling the oscillating field $\v E$ evoked by microwaves. Here one starts from the time derivative of the Ampère-Maxwell law in \eqref{eq:Maxwell},
\[ \rot{\partial_t \v B}=\mu_0 \partial_t \v J+\mu_0 \partial_{tt} \v D -\v S,\] and couples it with  Faraday's law $\partial_t \v B=-\rot{\v E}$ to get the
radiation equation
\begin{equation}\label{eq:Strahlung}\varepsilon\mu_0 \partial_{tt}\v E+\sigma\mu_0\partial_t \v E + \rot{\rot{\v E}}=\v S,
\end{equation} Here, $\v S$ is a known term, representing the induced electromagnetic field of the excitation. -- By vector analysis identities, \eqref{eq:Strahlung} is equivalent with 
\begin{equation}\label{eq:Strahlung}\varepsilon\mu_0 \partial_{tt}\v E+\sigma\mu_0\partial_t \v E + \grad{(\div{\v E})}-\gv\Delta\v E=\v S,
\end{equation}
where $\gv \Delta$ is the Laplace operator applied component-wise.

\subsubsection*{Imaging:}
The initial pressure distribution can be obtained by acoustic reconstruction formulas. To calculate the conductivity $\sigma$, we have now the quantitative aspect of thermo-acoustic tomography:
\begin{prob}[qTAT]\label{prob:qTAT}
Let
\[\varepsilon\mu_0 \partial_{tt}\v E+\sigma\mu_0\partial_t \v E + \grad{(\div{\v E})}-\gv\Delta\v E=\v S\]
with known $\v S$ and constant $\varepsilon$, $\mu_0$.

Determine the conductivity $\sigma$ from knowledge of $\v S$ and $\sigma|\v E|^2$.
\end{prob}
In \cite{BalRenUhlZho11}, unique recovery of $\sigma$ from $H=\sigma(\v x)|\v E(\v x)|^2$ is shown for a class of bounded conductivity distributions (assuming that $\varepsilon\equiv\text{const.}$, such that $\div{\v E}=0$ by the second eqn. in \eqref{eq:Maxwell}). Also, the following scalar model is investigated
\[ \varepsilon\mu_0\partial_{tt}u+\sigma\mu_0\partial_t u-\Delta u=S, \]
which is converted by the Fourier transform into
\[\Delta u+k^2u+ikqu=S.\]
Here, $k=\frac{1}{\sqrt{\varepsilon\mu_0}\omega}$ is the wave number, and $q=-\sigma\sqrt\frac{\mu_0}{\varepsilon}$.
An exact reconstruction formula for $q$ in this equation is developed in \cite{AmmGarJinNgu12} (also assuming $\varepsilon$ is constant).

\subsubsection{Magneto-acoustic tomography with magnetic induction (MAT-MI):}
\label{sec:MATMI}
Magneto-acoustic tomography exploits the Lorentz force effect to create a pressure wave.

\subsubsection*{Experiment:}
The measurement equipment consists of an excitation coil, a static magnet and ultrasound transducers. While the tissue is kept in the static magnetic field, electrical pulses in the excitation coil induce a eddy currents in the tissue. These, in turn, are affected by the Lorentz force due to the static field. The force causes the tissue to displace locally, and a pressure wave results. This wave reaches the boundary, where ultrasound transducers record the signal, either at the surface of the body or in a contact medium, such as water.

Experimental setups with different arrangements of induction coils are documented in \cite{XuHe05,HuLiHe10}.

\subsubsection*{Mathematical modeling:}
We model the ultrasound wave motion with the wave equation with outer force $\v f$. Applying \eqref{eq:Lorentz} with vanishing charge density $\rho=0$ to the static field $\v B=\v B_0$, we obtain the expression for the Lorentz force density $\v f=\v J\times \v B_0$ (see also \cite{RotWik98}). Following the discussion in section \ref{sec:AkustikTheorie}, the quantitative description of the acoustic wave is governed by

 \[
  \frac{1}{c^2} \partial_{tt} p(\v{x},t)- \Delta p(\v{x},t)  =-\div{( \v J\times\v B_0)}+\frac{\Gamma}{c^2}\partial_t H,\]
  where $H$ quantifies the absorbed energy due to eddy currents. Notwithstanding, the energy absorption effect can be ignored in the experiments of \cite{XuHe05}. There, magnetic induction is used for excitation of electrical currents, and this entails smaller field strengths. Considering the usual scale of conductivity in biological tissue, the pressure due to the thermo-acoustic effect can be neglected compared with pressure due to Lorentz force\footnote{Y. Xu, personal communication} (see also \cite[p.~5179]{XuHe05}).

  Therefore the model is
  \begin{equation}\label{DruckquellenMATMI}
    \frac{1}{c^2} \partial_{tt} p(\v{x},t)- \Delta p(\v{x},t)  =-\div{( \v J\times\v B_0)},
\end{equation}
together with the initial and boundary conditions
\begin{equation}\label{AnfangswerteMATMI}
p(0)=0,\qquad\partial_t p(0)=0.
\end{equation}

We assume that the time- and space variables can be separated. The inhomogeneity $\v f$ is therefore $\div{\v {\overline f}(\v{x})}g(t)$, where $\v{\overline  f}$ is the force density $-\v{J}\times \v{B_0}=-\sigma(\v{x})\v{\overline{E}}(\v{x})\times\v{B}_0(\v{x})$. Due to the different frequency ranges of the electrical and the acoustical pulse, we assume again that $g$ is a delta distribution. According to Duhamel's principle \eqref{eq:Duhamel},
the inhomogeneous problem  \eqref{DruckquellenMATMI} and \eqref{AnfangswerteMATMI} can be converted to a Cauchy problem for the homogeneous wave equation \cite{AmmCapKanKoz09, AmmKan11}:

\begin{equation}
\begin{aligned}
    \frac{1}{c^2}
\partial_{tt}p   -\Delta p&=0  && (\v{x},t) \in\Omega\times(0,\infty) \\
p&=0   && \v{x}\in\partial\Omega\\
\partial_t p&=-\div{(\sigma\v{E}\times\v{B}_0)} &&t=0 \\
p&=0 &&t=0. \\
\end{aligned}
\end{equation}

\subsubsection*{Imaging:}
The first step of the reconstruction of the source term $\div{(\sigma\v{E}\times\v{B}_0)}$. In contrast to the techniques of impedance-acoustic tomography and thermo-acoustic tomography, we now look for the initial velocity of the pressure, which can be recovered as well.

As quantitative problem, it remains to recover the distribution of the conductivity $\sigma$ over $\Omega$. For this purpose, one has to describe the electrical field $\v E$.
For this, we use the eddy current model \eqref{eq:IC-Poisson} discussed in section \ref{sec:BioWirbelstrom}, adapted here for the pulsed excitation current (see also \cite{LiMarHe10}). Therefore, the hybrid imaging problem is

\begin{prob}[Inverse problem MAT-MI]\label{prob:MATMI}
Let
\[\begin{aligned}
\div{(\sigma \grad{u})}&= -\partial_t\v{A_p}\cdot\grad{\sigma} \text{ in }\Omega \\
\partial_{\v n} u&=-\partial_t\v{ A_p}\cdot \v n \text{ on }\partial\Omega\\
\int_{\partial\Omega}u \;dS&=0.
\end{aligned}\]
The vector potential $\v{A_p}|_\Omega$ on $\mathbb{R}^3$ is determined from the excitation current current $\v{J_i}|_F$  by
\[
\v{A_p}(\v x,t)=\frac{\mu_0}{4\pi}\int_{\mathbb{R}^2} \frac{\v{J_i}(\v y,t)}{|\v{x}-\v{y}|} d^3\v{y}.
\]
Given are $\v B_0$ and $\div{(\sigma(\partial_t\v{A_p}+\grad{u})\times\v{B}_0)}$, from which to determine $\sigma$.

\end{prob}

\subsubsection{Magneto-acoustic-electrical tomography (MAET):}
\label{sec:MAET}

In MAET, pressure waves excite the tissue, and a voltage is produced due to the Lorentz force effect.

% Zitat Kunaynsky: Magneto-Acousto-Electric Tomography (MAET) is based on the measurements of the electrical potential arising when an acoustic wave propagates through conductive medium placed in a constant magnetic field. \cite{Kun11}

\subsubsection*{Experiment:}
The measurement equipment in MAET consists of a permanent magnet, a waveform generator, and of electrodes to measure the voltage at the boundary. The sample is put into the static magnetic field, and expopsed to different ultrasound waves. Due to frictional forces, the ions in the tissue move. As there is a magnetic field present, the electrical charges separate by the Lorentz force mechanism. A voltage results, which is detected on the surface of the probe. -- The experiment is repeated with different arrangements of the static field.

The physical effect underlying this technique has been studied in \cite{WenShaBal98,MonJosMatCat01}, further imaging experiments have been reported in \cite{HaiHrbXu2008}.

\subsubsection*{Modeling:}
To model MAET, we follow the approach in \cite{Kun11}. %Here, arbitrary sound waves are allowed. The more specific condition that focused ultrasound excitation is available was used for the modelling in \cite[9.6.1]{AmmKan11}. 
We assume that the static magnetic field $\v B$ in the experiments is homogeneous, that is, $\v B(\v x)=\v B$. The Lorentz force exerted upon the moving particles gives rise to an impressed current density $\v J_L$ which is approximately \cite{Kun11,MonJosMatCat01}
\[
\v J_L(\v x,t)=\sigma(\v x) \v B\times\v v(\v x,t),
\]
where $\v v$ is the velocity field of the fluid.

For each time $t$ the electrical field inside the domain $\Omega$ is described by an electrical potential, $\v E(\v x,t)=-\grad{u(\v x,t)}$. In absence of any other free charges, the potentials can be described by the conditions $\div{\v J}=0 $ and $\v J\cdot \v n=0$ for the total current $\v J=-\sigma \grad u+\v J_L$, that is
\begin{equation}\begin{aligned}
\div{(\sigma\grad{u})} & = -\div{(\sigma \v B\times \v v)},&\qquad\v x\text{ in }\Omega \\
\partial_{\v n} u 	   & = - (\v B\times\v v)\cdot \v n,  &\qquad\v x\text{ on }\partial\Omega
\end{aligned}\end{equation}

The voltage measured at the boundary gives knowledge of $u(\v x,t)|_{\partial\Omega}$. One can also assume that only a weighted average of $u$ is known, that is \begin{equation}\label{eq:KunMessung1}
M(t)=\int_{\partial\Omega}I(\v x) u(\v x,t) d S(\v x).
\end{equation}

\subsubsection*{Mathematical analysis:}
Using identities from vector analysis, it can be shown that the measurements in \eqref{eq:KunMessung1} are
\begin{equation}\label{eq:KunMessung2}
M(t)=-\v B\cdot\int_\Omega \v J_I\times \v v \;d^3\v x\,,
\end{equation}
with the current density $\v J_I(\v x)=-\sigma \grad{u_I}(\v x)$ satisfying the electrostatic equation
\begin{equation}\label{eq:KunNormalstrom}\begin{aligned}
\div{(\sigma\grad{u_I})} & = 0,&\qquad\v x\text{ in }\Omega \\
\sigma\partial_{\v n} u_I 	   & = I(\v x),&\qquad\v x\text{ on }\partial\Omega.
\end{aligned}\end{equation}
The inverse problem is
\begin{prob}[MAET]\label{prob:MAET}
Let $u_I$ satisfy \eqref{eq:KunNormalstrom}. Determine $\sigma$ from knowledge of $M(t)$ in \eqref{eq:KunMessung2}, by possibly varying $\v B$, $I$, and $\v v$.
\end{prob}

Reconstructions for the case of general acoustic waves have been given in \cite{Kun11}. The procedure leads to knowledge of several vector valued current densities $\v J_I$, which in the last step are converted to $\sigma$. This is a problem of current density impedance imaging (CDII), discussed in section \ref{sec:MREIT}. Reconstructions for the case of focused ultrasound excitation are given in \cite{AmmCapKanKoz09}.

\subsubsection{Acousto-electrical tomography (AET):}
\label{sec:AET}\bigskip
An EIT experiment is performed and, in addition, the conductivity values are perturbed via ultrasound excitation in parallel.

\subsubsection*{Experiment:}
For the measurements, one needs the typical EIT electrode setup and a device to produce ultrasound waves to excite the probe. The standard EIT measurement protocol is performed. For the perturbation of the conductivity, a set of various ultrasound waves is applied. The boundary voltages due to the perturbation are then compared with the standard voltages.

The method was proposed in \cite{ZhaWan04}. The underlying physical effect is treated in \cite{LavJosCat00}.

\subsubsection*{Modeling:}
The standard electrostatic equation from \eqref{eq:CalderonMaxwell} is used to model the electrical potential $u$,
\[\begin{aligned}
 \div{(\sigma \grad{u})}&=0 \qquad&&\text{in }\Omega \\
 \sigma \,\partial_\v n u&= g&&\text{on }\partial\Omega.
\end{aligned}\]
We now assume that the probe is excited by \emph{focused} ultrasound beams. Therefore, the conductivity (which depends on the pressure), is perturbed in a domain $D=\v z+\delta B$, where $B$ is the unit disc and $\v z$ is the center of focus. The perturbed conductivity is denoted $\sigma_{\v z}^\delta$; the new potential $u^\delta$ is therefore described by
\[\begin{aligned}
 \div{(\sigma_{\v z}^\delta \grad{u_{\v z}^\delta})}&=0 \qquad&&\text{in }\Omega \\
 \sigma^\delta_{\v z} \,\partial_\v n u_{\v z}^\delta&= g&&\text{on }\partial\Omega.
\end{aligned}\]
As measurement data, we have the difference of the measured voltages at the boundary, i.e. $u^\delta-u|_{\partial\Omega}$.

\subsubsection*{Mathematical analysis:}
It is shown in \cite{AmmBonCapTanFin08}, that for $\delta\to 0$, one has the asymptotic formula
\[\int_{\partial \Omega}(u^\delta-u) \, g\, dS = |\grad{u} (\v z)|^2 \int_D\sigma(\v y)\frac{(\eta(\v{y})-1)^2}   {\eta(\v{y})+1} d^3\v{y} + o(\operatorname{vol}(D)).\]
Here, $\eta$ is a known function in $\Omega$.

From this formulation, one notices that the measurement data give knowledge of the power density at $\v z$, $\sigma(\v z) |\v \grad{u}(\v z)|^2$, up to a small error.

\subsubsection*{Imaging:}
The inverse problem is to recover the conductivity $\sigma$ from the interior information $\sigma|\grad{u}|$. This problem is identical with problem \ref{prob:IAT} of \emph{impedance-acoustic tomography}.

\bigskip

\begin{prob}[AET, conductivity determination]\label{prob:AET}
Let
\[\begin{aligned}\div{(\sigma \grad{u})}&=0 \qquad	&& \text{in } \Omega\\
u|_{\partial\Omega}&= f && \text{on } \partial\Omega\\
\sigma\,\partial_\v nu&= g && \text{on } \partial\Omega.\end{aligned}\]
Given are $f$, $g$, as well as interior data $H=\sigma|\v \grad{u}|^2$. Determine $\sigma$.
\end{prob}

\subsubsection*{Mathematical analysis:}
In the modeling of AET, we have assumed that $\sigma_{\v z}^\delta-\sigma$ are approximately delta-distributions. But this restriction can be overcome and other more practical forms of ultrasound perturbation (such as plane waves, e.g.) can be used \cite{KucKun10}, \cite[Section 2]{KucKun11b}.

For this we consider
\[
L_\sigma(f)(\v y)=\int_\Omega k(\v x,\v y) f(\v x)\; d^3 \v x
\]
where $L_\sigma$ maps an increment $f=d\sigma$ of conductivity to the corresponding perturbation $L_\sigma f$ in the boundary voltage. %%% Etwas zu einer Linearisierung sagen.
As perfect focusing of ultrasound waves correspond to $\delta$-distributions $f$, and this would correspond to simple evaluation of the kernel $k(x,y)$. It is more realistic that  ultrasound waves of other, unfocused type, trigger the corresponding conductivity changes, proportional to the pressure. So, the intermediate step (\emph{synthetic focusing}) corresponds to finding the values of the kernel $k(x,y)$ by measurements of
\[
\int_\Omega k(\v x,\v y) w_\alpha(\v x) d^3{\v x}
\]
for a particular class of waves $w_\alpha$. Synthetic focusing for spherical waves, plane waves has been discussed in \cite[Section 2]{KucKun10}; numerical calculations are performed in \cite{KucKun11b}.

\section{Summary}
The hybrid imaging techniques and mathematical models discussed above are summarized in the following tables \ref{tab:quantitative1}, \ref{tab:quantitative2}. Those, consist of four columns, summarizing the imaging data, the overspecified data of the quantitative inverse problem, and the equations of the forward problems. In mathematical terms, in general, these are inverse problems with interior data. The table, however, does not summarize unique identifability and unique reconstructability of the quantitative imaging data. Hybrid models coupling magnetic resonance effects with current densities are listed in table \ref{tab:quantitative1}. Some of these, like MREIT/CDII, rely on the low-frequency description $\div{(\sigma\grad{u})}=0$, others use radiofrequency currents, where the electrostatic approximation is not available and quasistatic approximations are used. Coupling hybrid techniques with ultrasound modalities are summarized in table \ref{tab:quantitative2}. They also range from low-frequency or quasistatic to high-frequency ranges.\footnote{Shortly, we also mention some hybrid imaging techniques using other physical parameters, which nevertheless are formally related to the models using the electrostatic regime, $\div{(\sigma\grad{u})}=0$:
\begin{description}
\item {In \emph{quantitative photoacoustic tomography} (qPAT)} in the diffusion regime, the diffusivity $\kappa$ is reconstructed, while the absorption $\mu$ in the elliptic equation $\div({\kappa \grad{u}})+\mu u=0$ is known. The imaging data are $H=\mu I$ \cite{BalUhl10,CoxLauBea09}.
\item In \emph{microwave tomography}, the generalization of AET, one recovers the parameters $a$ and $q$ in $\div{(a\grad{u})}+k^2\, q\, u=0$, with $a|\grad{u}|^2$ and $q|u|$ given \cite{AmmKan11,AmmCapGouRozTri11}.
\item In \emph{pptoacoustic tomography} one also recovers $\sigma$ in $\div{(\sigma\grad{u})}=0$ with $\sigma|\grad{u}|^2$ given, where $\sigma$ is the optical absorption \cite{BalSch10}.
\end{description}}

In addition to coupled Physics (hybrid) techniques with interior information, we also summarize the standard EIT problem (the Calderón problem) together with techniques using electromagnetic coupling in several frequency ranges (see tables \ref{tab:eit}). The corresponding reconstruction cannot be separated into two imaging problems with interior data.

\begin{table}

% Tabellenbeschriftung oberhalb der Tabelle gemaess IOP Richtlinie
\caption{\label{tab:eit} Electromagnetic variants of the Calderón problem in different frequency ranges.}

\begin{tabular}{|c|x{.29\textwidth}|x{.2\textwidth}|x{.22\textwidth}|}
\hline
Technique & equation in $\Omega$ & boundary data & remarks \\
\hline
\hline
%% CALDERON
   $\begin{array}{c}\text{EIT}\\ \text{eqn. }\eqref{Calderon}
   \end{array}$ & 
   $\div{  \sigma \, \grad{u} }=0$	&	
   $ \begin{array}{@{}l@{}} u \\ \sigma\,\partial_\v n u \end{array} $ & \\
\hline
\hline
%%% TOZER
$\begin{array}{c} \text{IC-EIT} \\
                  \text{section } \ref{sec:ICEIT}
                  \end{array}$ &
    $\div{\left (\sigma \grad{u}\right )}=-i\omega \v{A}  \cdot\grad{\sigma}$	&	
    $\begin{array}{@{}l@{}} u \\ \sigma\,\partial_\v n u=-i\omega \v{A}\cdot\v{n} \end{array}$ & 
    $\begin{array}{@{}l@{}} \text{excitation } \v A  \text{ known} \end{array}$\\
\hline
%% SCHARFETTER
           
                        $\begin{array}{c} \text{MIT} \\ \text{section  }\ref{sec:MIT} \end{array}$ & 
    $\div{\left (\sigma \grad{u}\right )}=-i\omega \v{A}  \cdot\grad{\sigma}$
    &
    $\begin{array}{@{}l@{}}\int_C \v n\cdot\v{B}\\ \sigma\,\partial_\v n u=-i\omega \v{A}\cdot\v{n}\end{array}$
    &
    $\begin{array}{@{}l@{}} \text{excitation } \v A \text{ known}. \\
                       \v B=\v B(\sigma,u,\v A),\\
                        C \text{ is a coil surface }\\ \text{outside $\Omega$}.\end{array}$\\

\hline
\hline
%%% AHLFORS
$\begin{array}{c} \text{MD-EIT}\\ \text{section } \ref{sec:MDEIT}	\end{array}$ &	$\begin{array}{@{}l@{}} \v{B}=\int \frac{\v{J}\times \v{R}}{|\v{R}|^3}\end{array}$	&  $\v{B}$  & $\begin{array}{@{}l@{}}\text{recover }\v J\text{, not }\sigma \end{array}$ \\
\hline

\end{tabular}

\end{table}

\begin{table}
%\begin{sidewaystable}\centering

% Tabellenbeschriftung oberhalb der Tabelle gemaess IOP Richtlinie
\caption{\label{tab:quantitative1} Coupled Physics models with interioir data, based on MRI.}

%% Wenn man \textwidth statt cm als Referenz waehlt, dann wechseln beim Rotieren der Tabelle alle Spaltenbreiten
\begin{tabular}{|c|x{2.3cm}|x{4.0cm}|x{3.2cm}|x{2.6cm}|}

\hline
Technique & interior data&equation in $\Omega$ & boundary data&remarks\\
\hline
%\hline
%%% NACHMAN
% $\begin{array}{c} \text{CDII} \\
%                  \text{section } \ref{sec:MREIT}
%                  \end{array}$ 	&$\v J_1$, $\v J_2$\newline $\sigma(\v x_0)$&	$\grad{\log (\sigma)}=G(\v J_1,\v J_2)$	&		 & \\
%\hline
%
%$\begin{array}{c} \text{CDII} \\
%                  \text{section } \ref{sec:MREIT}
%                  \end{array}$&$\v J=-\sigma\grad{u}$&$\div{(\sigma\grad{u})}=0$&$u$& first recover $u$, then $\sigma=|\v J|/|\grad{u}| $\\
%\hline
%
%	$\begin{array}{c} \text{CDII} \\
%                  \text{section } \ref{sec:MREIT}
%                  \end{array}$& $|\v J_1|$, $|\v J_2|$& $\div{ \left ( \sigma \, \grad u_i \right )}=0$	&	 $\sigma\,\partial_\v n u_i$ &  		 first recover $u_i$ then \newline $\sigma=|\v{J}_i|/|\grad u_i|$    	\\
%\hline
%	
	
	$\begin{array}{c} \text{CDII} \\
                  \text{section } \ref{sec:MREIT}
                  \end{array}$&$\begin{array}{l}\sigma\,\grad{u}\text{ or }\\ \sigma|\grad{u}|\end{array}$& $\div{ \left ( \sigma \, \grad u \right )}=0$	&	$u$ &  		    	\\
\hline

%% SEO, KWON
 $\begin{array}{c} \text{MREIT} \\
                  \text{section } \ref{sec:MREIT}
                  \end{array}$&$B^z$&$\div{ \left ( \sigma \, \grad u \right )}=0$	&  $ \sigma\,\partial_\v n u$	&	 $B^z=F(\sigma \grad{u},\sigma)$  \\
\hline

%% HE: ICMREIT
$\begin{array}{c} \text{CDII-MI} \\
                  \text{section } \ref{sec:ICMREIT}
                  \end{array}$&$\sigma\v E$&	$\div{\left (\sigma \grad{u}\right )}=-i\omega \v{A}  \cdot\grad{\sigma}$&   $\sigma\,\partial_\v n u=i\omega \v{A}  \cdot\v{n}$    &excitation $\v A$ known \newline $\v E=-i\omega\v A-\grad{u}$ \\
\hline

$\begin{array}{c} \text{ICMREIT} \\
                  \text{section } \ref{sec:ICMREIT}
                  \end{array}$&$B^z$&	$\div{\left (\sigma \grad{u}\right )}=-i\omega \v{A}  \cdot\grad{\sigma}$&   $\sigma\,\partial_\v n u=i\omega \v{A}  \cdot\v{n}$    &excitation $\v A$ known \newline $B^z=F(\sigma \grad{u},\sigma)$ \\

\hline
\end{tabular}

%\end{table}
%\end{sidewaystable}

%\newpage\thispagestyle{plain}
%\begin{table}
%\begin{sidewaystable}\centering

\vspace{1cm}

% Tabellenbeschriftung oberhalb der Tabelle gemaess IOP Richtlinie
\caption{\label{tab:quantitative2} Coupled Physics models with interior data, based on Ultrasound.}

%% Wenn man \textwidth statt cm als Referenz waehlt, dann wechseln beim Rotieren der Tabelle alle Spaltenbreiten
\begin{tabular}{|c|x{2.3cm}|x{4.0cm}|x{3.2cm}|x{2.6cm}|}

\hline
Technique & interior data &equation in $\Omega$ & boundary data &remarks\\
\hline
\hline

%% GEBAUER-SCHERZER
$\begin{array}{c} \text{IAT} \\
                  \text{section } \ref{sec:IAT}
                  \end{array}$
&	$\sigma|\grad{u}|^2$&$\div{(\sigma\grad{u})}=0 	$&$u$\newline $\sigma\partial_{\v n} u$	&\\
\hline
%% WANG ??
$\begin{array}{c} \text{qTAT} \\
                  \text{section } \ref{sec:qTAT}
                  \end{array}$
& $\sigma|\grad{u}|^2$	& $\varepsilon\mu_0 \partial_{tt}\v E+\sigma\mu_0\partial_t \v E +$\newline $ \rot{\rot{\v E}}=\v S$	 &$\v S|_F$	& $F\cap\Omega=\emptyset$		\\
\hline

$\begin{array}{c} \text{MAT-MI} \\
                  \text{section } \ref{sec:MATMI}
                  \end{array}$
&$ \div{(\sigma \v E\times \v B_0)}$&
$\div{\left (\sigma \grad{u}\right )}=-\partial_t \v{A}  \cdot\grad{\sigma}$
&$\sigma\,\partial_\v n u=-\partial_t \v{A}  \cdot\v{n}$&
excitation $\v A$ known, \newline $\v E=-\partial_t\v A-\grad{u}$ \\

%% BIN HE
%MAT-MI	&\ref{sec:MATMI}& $ \div{(\sigma \v E\times \v B_0)}$& $\sigma\mu_0\partial_t %\v E + \rot{\rot{\v E}}=\v{J_i}$ &	 $\v{J}_i|_F$	& $F\cap\Omega=\emptyset$\newline %$\v{B}_0|_{\mathbb{R}^3}$ known\vspace{1.5cm} \\
\hline
%% ROTH
%Montalibet 2001, \newline Roth 2009	&	VPT (Hall-Effekt-Verfahren)	&	$L(\sigma,p,I_0,u_end)=0$	&$u$ &	$\v{E}=\grad{u}$	 \\
%\hline
%% AMMARI
$\begin{array}{c} \text{MAET} \\
                  \text{section } \ref{sec:MAET}
                  \end{array}$
 & $\v B\cdot\int \sigma\grad{u_I}\times \v v$&$\div{(\sigma\grad{u_I})}=0 	$& $u_I$\newline $\sigma\partial_{\v n} u_I 	    = I$&
$\v B\cdot\int_\Omega \sigma\grad{u_I}\times \v v =\int_{\partial\Omega}I\, u\, d S$\\
\hline
$\begin{array}{c} \text{AET} \\
                  \text{section } \ref{sec:AET}
                  \end{array}$
	&	$\sigma|\grad{u}|^2$&$\div{(\sigma\grad{u})}=0 	$&$u$\newline $\sigma\partial_{\v n} u$	&\\
\hline

\end{tabular}

\end{table}
%\end{sidewaystable}

\bigskip

\section{Acknowledgements}
We thank Barbara Kaltenbacher for helpful discussions and Yuan Xu, Eung Je Woo, Doga G\"ursoy and Hermann Scharfetter for information concerning the special techniques.
This work has been supported by the Austrian Science Fund (FWF)
within the national research network Photo\-acoustic Imaging in Biology and Medicine,
project S10505-N20.

\bigskip

%\nocite{Wid11}
%\bibliographystyle{\BibPath plain_talks}
%\bibliography{\BibPath strings,\BibPath articles,\BibPath books,\BibPath infmath,\BibPath infmath_books,\BibPath infmath_report,\BibPath infmath_talks,\BibPath infmath_theses,\BibPath inproceedings,\BibPath preprints,\BibPath proceedings,\BibPath theses,\BibPath unsubmitted,\BibPath websites}

\begin{thebibliography}{100}

\bibitem{AdlGabLio11}
A.~Adler, R.~Gaburro, and W.~Lionheart.
\newblock Electrical impedance tomography.
\newblock In O.~Scherzer, editor, {\em \cite{Sch11}}, pages 599--652. Springer,
  2011.

\bibitem{Aiz10}
A.~P. Aizebeokhai.
\newblock {2D} and {3D} geoelectrical resistivity imaging: theory and field
  design.
\newblock {\em Sci. Res. Essays}, 23:3592--3605, 2010.

\bibitem{Ale88}
G.~Alessandrini.
\newblock Stable determination of conductivity by boundary measurements.
\newblock {\em Appl. Anal.}, 27:153--172, 1988.

\bibitem{Amm08}
H.~Ammari.
\newblock {\em An Introduction to Mathematics of Emerging Biomedical Imaging}.
\newblock Math{\'e}matiques \& Applications. Springer, 2008.

\bibitem{AmmBonCapTanFin08}
H.~Ammari, E.~Bonnetier, Y.~Capdeboscq, M.~Tanter, and M.~Fink.
\newblock Electrical impedance tomography by elastic deformation.
\newblock {\em SIAM J. Appl. Math.}, 68(6):1557--1573, 2008.

\bibitem{AmmCapGouRozTri11}
H.~Ammari, Y.~Capdeboscq, F.~de~Gournay, A.~Rozanova, and F.~Triki.
\newblock Microwave imaging by elastic perturbation.
\newblock {\em SIAM J. Appl. Math.}, 71:2112--2130, 2011.

\bibitem{AmmCapKanKoz09}
H.~Ammari, Y.~Capdeboscq, H.~Kang, and A.~Kozhemyak.
\newblock Mathematical models and reconstruction methods in magneto-acoustic
  imaging.
\newblock {\em European J. Appl. Math.}, 20(3):303--317, 2009.

\bibitem{AmmGarJinNgu12}
H.~Ammari, J.~Garnier, W.~Jing, and L.~Nguyen.
\newblock Quantitative thermo-acoustic imaging: and exact reconstruction
  formula.
\newblock Preprint, 2012.
\newblock http://arxiv.org/abs/1201.0619v1.

\bibitem{AmmKan11}
H.~Ammari and H.~Kang.
\newblock Expansion methods.
\newblock In O.~Scherzer, editor, {\em \cite{Sch11}}, pages 447--500. Springer,
  2011.

\bibitem{ArfWeb01}
G.~B. Arfken and H.~J. Weber.
\newblock {\em Mathematical Methods for Physicists}.
\newblock Academic Press, 5th edition, 2001.

\bibitem{AstPai06}
K.~Astala and L.~P{\"a}iv{\"a}rinta.
\newblock Calder{\'o}n's inverse conductivity problem in the plane.
\newblock {\em Ann. Math.}, 163:265--99, 2006.

\bibitem{Bal11a}
G.~Bal.
\newblock Cauchy problem for ultrasound modulated {EIT}.
\newblock Preprint, 2011.

\bibitem{Bal11b}
G.~Bal.
\newblock Hybrid inverse problems and internal information.
\newblock Preprint, 2011.
\newblock http://arxiv.org/abs/1110.4733v1.

\bibitem{BalBonMonTri11}
G.~Bal, E.~Bonnetier, F.~Monard, and F.~Triki.
\newblock Inverse diffusion from knowledge of power densities.
\newblock Preprint, 2011.
\newblock http://arxiv.org/abs/1110.4577v1.

\bibitem{BalRenUhlZho11}
G.~Bal, K.~Ren, G.~Uhlmann, and T.~Zhou.
\newblock Quantitative thermo-acoustics and related problems.
\newblock {\em Inverse Probl.}, 27:055007, 2011.

\bibitem{BalSch10}
G.~Bal and J.~C. Schotland.
\newblock Inverse scattering and acousto-optic imaging.
\newblock {\em Phys. Rev. Lett.}, 104:043902, 2010.

\bibitem{BalUhl10}
G.~Bal and G.~Uhlmann.
\newblock Inverse diffusion theory of photoacoustics.
\newblock {\em Inverse Probl.}, 26:085010, 2010.

\bibitem{Bay06}
R.~H. Bayford.
\newblock Bioimpedance tomography (electrical impedance tomography).
\newblock {\em Annu. Rev. Biomed. Eng.}, 8:63--91, 2006.

\bibitem{BinLyoPeyPri01}
R.~Binns, A.~R.~A. Lyons, A.~J. Peyton, and W.~D.~N. Pritchard.
\newblock Imaging molten steel flow profiles.
\newblock {\em Meas. Sci. Technol.}, 12:1132--1138, 2001.

\bibitem{Bor02}
L.~Borcea.
\newblock Electrical impedance tomography.
\newblock {\em Inverse Probl.}, 18:R99--R136, 2002.

\bibitem{Cal80}
A.~P. Calder{\'o}n.
\newblock On an inverse boundary value problem.
\newblock In {\em Seminar on numerical analysis and its applications to
  continuum physics}, pages 65--73. Soc. Brasil. Mat., Rio de Janeiro, 1980.

\bibitem{CapFehGouKav09}
Y.~Capdeboscq, J.~Fehrenbach, F.~de~Gournay, and O.~Kavian.
\newblock Imaging by modification: numerical reconstruction of local
  conductivities from corresponding power density measurements.
\newblock {\em SIAM J. Imaging Sciences}, 2(4):1003--1030, 2009.

\bibitem{CheIsaNew99}
M.~Cheney, D.~Isaacson, and J.~C. Newell.
\newblock Electrical impedance tomography.
\newblock {\em SIAM Rev.}, 41(1):85--101, 1999.

\bibitem{CoxLauBea09}
B.~T. Cox, J.~G. Laufer, and P.~C. Beard.
\newblock The challenges for quantitative photoacoustic imaging.
\newblock {\em Proc. SPIE}, 7177:717713, 2009.

\bibitem{DekYinKtiArmPey10}
B.~Dekdouk, W.~Yin, Ch. Ktistis, D.~W. Armitage, and A.~J. Peyton.
\newblock A method to solve the forward problem in magnetic induction
  tomography based on the weakly coupled field approximation.
\newblock {\em IEEE Trans. Biomed. Eng.}, 57(4):914--921, 2010.

\bibitem{DubPiwSanWee78}
A.~Duba, A.~J. Piwinskii, M.~Santor, and H.~C. Weed.
\newblock The electrical conductivitiy of sandstone, limestone and granite.
\newblock {\em Geophys. J. Int.}, 53:583--597, 1978.

\bibitem{EftSamPol10}
K.~G. Efthimiadis, T.~Samaras, and K.~S. Polyzoidis.
\newblock Magnetic stimulation of the spine: the role of tissues and their
  modeling.
\newblock {\em Phys. Med. Biol.}, 55:2541--2553, 2010.

\bibitem{Emi92}
C.~Emiliani.
\newblock {\em Planet Earth. Cosmology, Geology, and the Evolution of Life and
  Environment}.
\newblock Cambridge University Press, 1992.

\bibitem{EngSte11}
S.~Engleder and O.~Steinbach.
\newblock Boundary integral formulations for the forward problem in magnetic
  induction tomography.
\newblock {\em Math. Methods Appl. Sci.}, pages 1144--1156, 2011.

\bibitem{Eva98}
L.~C. Evans.
\newblock {\em Partial Differential Equations}, volume~19 of {\em Graduate
  Studies in Mathematics}.
\newblock American Mathematical Society, Providence, RI, 1998.

\bibitem{GabLauGab96}
S.~Gabriel, R.~W. Lau, and C.~Gabriel.
\newblock The dielectric properties of biological tissues: {II}. measurements
  in the frequency range 10 {Hz} to 20 {GHz}.
\newblock {\em Phys. Med. Biol.}, 41:2251--69, 1996.

\bibitem{GebSch08}
B.~Gebauer and O.~Scherzer.
\newblock Impedance-acoustic tomography.
\newblock {\em SIAM J. Appl. Math.}, 69(2):565--576, 2008.

\bibitem{GeeCreDup10}
N.~De Geeter., G.~Crevecoeur, and L.~Dupr{\'e}.
\newblock Low-parametric induced current-magnetic resonance electrical
  impedance tomography for quantitative conductivity estimation of brain
  tissues using a priori information: a simulation study.
\newblock In {\em Engineering in Medicine and Biology Society (EMBC), 2010
  Annual International Conference of the IEEE}, pages 5669--5672, Buenes Aires,
  2010.

\bibitem{GenKozIde94}
N.~G. Gen\c{c}er, M.~Kozuoglu, and Y.~Z. \.{I}der.
\newblock Electrical impedance tomography using induced currents.
\newblock {\em IEEE Trans. Med. Imag.}, 13(2):338--350, 1994.

\bibitem{GurMamAdlScha11}
D.~G{\"u}rsoy, Y.~Mamatjan, A.~Adler, and H.~Scharfetter.
\newblock Enhancing impedance imaging through multimodal tomography.
\newblock {\em IEEE Trans. Biomed. Eng.}, 58(11):3215--3224, 2011.

\bibitem{GurScha09}
D.~G{\"u}rsoy and H.~Scharfetter.
\newblock Optimum receiver array design for magnetic induction tomography.
\newblock {\em IEEE Trans. Biomed. Eng.}, 56(5):1435--41, 2009.

\bibitem{HaaBroThoVen99}
E.~M. Haacke, R.~W. Brown, M.~R. Thompson, and R.~Venkatesan.
\newblock {\em Magnetic Resonance Imaging: Physical Principles and Sequence
  Design}.
\newblock Wiley, New York, 1999.

\bibitem{HaeStaTsaTunMah03}
D.~Haemmerich, S.~T. Stalin, J.~Z. Tsai, S.~Tungjitkusolmun, D.~M. Mahvi, and
  J.~G. Webster.
\newblock In vivo electrical conductivitiy of hepatic tumours.
\newblock {\em Physiol. Meas.}, 24:251--260, 2003.

\bibitem{HaiHrbXu2008}
S.~Haider, A.~Hrbek, and Y.~Xu.
\newblock Magneto-acousto-electrical tomography: a potential method for imaging
  current density and electrical impedance.
\newblock {\em Physiol. Meas.}, 29:S41--S50, 2008.

\bibitem{Hal10_habil}
M.~Haltmeier.
\newblock {\em Mathematical Methods in Photoacoustic Image Reconstruction}.
\newblock Habilitation, University of Vienna, Austria, Vienna, February 2010.

\bibitem{HasMaNacJoy08}
K.~F. Hasanov, A.~W. Ma, A.~I. Nachman, and M.~L.~G. Joy.
\newblock Current density impedance imaging.
\newblock {\em IEEE Trans. Med. Imag.}, 27(9):1301--1309, 2008.

\bibitem{HauKuhPot05}
K.-H. Hauer, L.~K{\"u}hn, and R.~Potthast.
\newblock On uniqueness and non-uniqueness for current reconstruction from
  magnetic fields.
\newblock {\em Inverse Probl.}, 21(3):955--967, 2005.

\bibitem{HauPotWan08}
K.-H. Hauer, R.~Potthast, and M.~Wannert.
\newblock Algorithms for magnetic tomography -- on the role of \emph{a priori}
  knowledge and constraints.
\newblock {\em Inverse Probl.}, 24(4):045008, 2008.

\bibitem{HuLiHe10}
G.~Hu, X.~Li, and B.~He.
\newblock Imaging biological tissues with electrical conductivity contrast
  below 1 ${S}\,m^{-1}$ by means of magnetoacoustic tomography with magnetic
  induction.
\newblock {\em Appl. Phys. Lett.}, 97:103705, 2010.

\bibitem{Hum30}
J.~N. Hummel.
\newblock Der scheinbare spezifische {W}iderstand.
\newblock {\em Z. Geophysik}, 5:89--104, 1930.

\bibitem{IreTozBarBar04}
R.~H. Ireland, J.~C. Tozer, A.~T. Barker, and D.~C. Barber.
\newblock Towards magnetic detection electrical impedance tomography: data
  acquisition and image reconstruction of current density in phantoms and in
  vivo.
\newblock {\em Physiol. Meas.}, 25:775--796, 2004.

\bibitem{Jac98}
J.~D. Jackson.
\newblock {\em Classical Electrodynamics}.
\newblock Wiley, 3rd edition, 1998.

\bibitem{JoiZhaLiJir94}
W.~Joines, Y.~Zhang, C.~Li, and Jirtle R.
\newblock The measured electrical properties of normal and malignant human
  tissues from 50 to 900 {MHz}.
\newblock {\em Med. Phys.}, 21:547--550, 1994.

\bibitem{Jos98}
J.~Jossinet.
\newblock The impedivity of freshly excised human breast tissue.
\newblock {\em Physiol. Meas.}, 19:61--75, 1998.

\bibitem{KatDorFinVerNeh07}
U.~Katscher, T.~Dorniok, C.~Findeklee, P.~Vernickel, and K.~Nehrke.
\newblock In vivo determination of electric conductivity and permittivity using
  a standard mr system.
\newblock In H.~Scharfetter and R.~Merwa, editors, {\em 13th International
  Conference on Electrical Bioimpedance and 8th Conference on Electrical
  Impedance Tomography 2007}, IFMBE Proceedings Vol. 17, pages 508--511.
  Springer, 2007.

\bibitem{KatVoiFinVerNeh09}
U.~Katscher, T.~Voigt, Ch. Findeklee, P.~Vernickel, K.~Nehrke, and
  O.~D{\"o}ssel.
\newblock Determination of electric conductivitiy and local {SAR} via {B1}
  mapping.
\newblock {\em IEEE Trans. Med. Imag.}, 28:1365--1374, 2009.

\bibitem{KesKesSma06}
A.~Keshtkar, A.~Keshtkar, and R.~H. Smallwood.
\newblock Electrical impedance spectroscopy and the diagnosis of bladder
  pathology.
\newblock {\em Physiol. Meas.}, 27:585--596, 2006.

\bibitem{KimKwoSeoYoo02}
S.~Kim, O.~Kwon, J.~K. Seo, and J.-R. Yoon.
\newblock On a nonlinear partial differential equation arising in magnetic
  resonance impedance tomography.
\newblock {\em SIAM J. Math. Anal.}, 34(3):511--526, 2002.

\bibitem{KimKwoSeoWoo03}
Y.~J. Kim, O.~Kwon, J.~Keun Seo, and E.~J. Woo.
\newblock Uniqueness and convergence of conductivity imge reconstruction in
  magnetic resonance electrical impedance tomography.
\newblock {\em Inverse Probl.}, 19:1213--1225, 2003.

\bibitem{KreKuhPot02}
R.~Kress, L.~K{\"u}hn, and R.~Potthast.
\newblock Reconstruction of a current distribution from its magnetic field.
\newblock {\em Inverse Probl.}, 18:1127--1146, 2002.

\bibitem{Kuc11}
P.~Kuchment.
\newblock Mathematics of hybrid imaging. {A} brief review.
\newblock Preprint, 2011.
\newblock http://arxiv.org/abs/1107.2447v1.

\bibitem{KucKun10}
P.~Kuchment and L.~Kunyansky.
\newblock Synthetic focusing in ultrasound modulated tomography.
\newblock {\em Inverse Probl. Imaging}, 4(4):665--673, 2010.

\bibitem{KucKun11b}
P.~Kuchment and L.~Kunyansky.
\newblock {2D} and {3D} reconstructions in acousto-electric tomography.
\newblock {\em Inverse Probl.}, 27(5):055013, 2011.

\bibitem{KucKun11}
P.~Kuchment and L.~Kunyansky.
\newblock Mathematics of photoacoustic and thermoacoustic tomography.
\newblock In O.~Scherzer, editor, {\em \cite{Sch11}}, pages 817--867. Springer,
  2011.

\bibitem{KucSte11}
P.~Kuchment and D.~Steinhauer.
\newblock Stabilizing inverse problems by internal data.
\newblock Preprint, 2011.
\newblock http://arxiv.org/abs/1110.1819v2.

\bibitem{Kun11}
L.~Kunyansky.
\newblock A mathematical model and inversion procedure for
  magneto-acousto-eletric tomography ({MAET}).
\newblock Preprint, 2011.
\newblock http://arxiv.org/abs/1108.0376v1.

\bibitem{KwoLeeYoo02}
O.~Kwon, J.-Y. Lee, and J.-R. Yoon.
\newblock Equipotential line method for magnetic resonance electrical impedance
  tomography.
\newblock {\em Inverse Probl.}, 18:1089--1100, 2002.

\bibitem{KwoWooYooSeo02}
O.~Kwon, E.~J. Woo, J.-R. Yoon, and J.~K. Seo.
\newblock Magnetic resonance electrical impedance tomography ({MREIT}):
  simulation study of {$J$}-substitution algorithm.
\newblock {\em IEEE Trans. Biomed. Eng.}, 48:160--167, 2002.

\bibitem{LanLif87}
L.D. Landau and E.M. Lifschitz.
\newblock {\em Course of Theoretical Physics, Volume 6: Fluid Mechanics}.
\newblock Pergamon Press, New York, 2nd edition, 1987.

\bibitem{LauIvoReuRubSol10}
S.~Laufer, A.~Ivorra, V.~E Reuter, B.~Rubinsky, and S.~B Solomon.
\newblock Electrical impedance characterization of normal and cancerous human
  hepatic tissue.
\newblock {\em Physiol. Meas.}, 31:995--1009, 2010.

\bibitem{LavJosCat00}
B.~Lavandier, J.~Jossinet, and D.~Cathignol.
\newblock Quantitative assessment of ultrasound-induced resistance change in
  saline solution.
\newblock {\em Med. Biol. Eng. Comput.}, 38:150--155, 2000.

\bibitem{Lee04}
J.-Y. Lee.
\newblock A reconstruction formula and uniqueness of conductivity in mreit
  using two internal current distributions.
\newblock {\em Inverse Probl.}, 20(3):847--858, 2004.

\bibitem{Leh08}
G.~Lehner.
\newblock {\em Elektromagnetische Feldtheorie}.
\newblock Springer, 6th edition, 2008.

\bibitem{LiWan09}
C.~Li and L.~V. Wang.
\newblock Photoacoustic tomography and sensing in biomedicine.
\newblock {\em Phys. Med. Biol.}, 54:R59--R97, 2009.

\bibitem{LiMarHe10}
X.~Li, L.~Mariappan, and B.~He.
\newblock Three-dimensional multiexcitation magnetoacoustic tomography with
  magnetic induction.
\newblock {\em J. App. Phys.}, 108:124702, 2010.

\bibitem{LiuSeoSinWoo07}
J.~J. Liu, J.~K. Seo, M.~Sini, and E.~J. Woo.
\newblock On the convergence of the harmonic ${B}_z$ algorithm in magnetic
  resonance electrical impedance tomography.
\newblock {\em SIAM J. Appl. Math.}, 67:1259--82, 2007.

\bibitem{LiuZhoHe09}
Y.~Liu, S.~Zho, and B.~He.
\newblock Induced current magnetic resonance electrical impedance tomography of
  brain tissues based on the ${J}$-substitution algorithm: a simulation study.
\newblock {\em Phys. Med. Biol.}, 54:4561--4573, 2009.

\bibitem{LusReiSte09}
H.~Lustfeld, M.~Rei{\ss}el, and B.~Steffen.
\newblock Magnetotomography and electric currents in a fuel cell.
\newblock {\em Fuel Cells}, 9(4):474--4781, 2009.

\bibitem{Man01}
N.~Mandache.
\newblock Exponential instability in an inverse problem for the schr{\"o}dinger
  equation.
\newblock {\em Inverse Probl.}, 17:1435--1444, 2001.

\bibitem{MikPavHar06}
D.~Miklavicic, N.~Pavselj, and F.~X. Hart.
\newblock Electric properties of tissues.
\newblock In {\em Wiley Encyclopedia of Biomedical Engineering}, pages 1--12.
  John Wiley \& Sons, 2006.

\bibitem{MonJosMatCat01}
A.~Montalibet, J.~Jossinet, A.~Matias, and D.~Cathignol.
\newblock Electric current generated by ultrasonically induced {Lorentz} force
  in biological media.
\newblock {\em Med. Biol. Eng. Comput.}, 39(1):15--20, 2001.

\bibitem{MorKun62}
J.~H. Moran and K.~S. Kunz.
\newblock Basic theory of induction logging and application to study of
  two-coil sondes.
\newblock {\em Geophysics}, 27(6):829--858, 1962.

\bibitem{NacTamTim07}
A.~Nachman, A.~Tamasan, and A.~Timonov.
\newblock Conductivity imaging with a single measurement of boundary and
  interior data.
\newblock {\em Inverse Probl.}, 23(6):2551--2563, 2007.

\bibitem{NacTamTim09}
A.~Nachman, A.~Tamasan, and A.~Timonov.
\newblock Recovering the conductivity from a single measurement of interior
  data.
\newblock {\em Inverse Probl.}, 25(3):035014, 2009.

\bibitem{NacTamTim10}
A.~Nachman, A.~Tamasan, and A.~Timonov.
\newblock Reconstruction of planar conductivities in subdomains from incomplete
  data.
\newblock {\em SIAM J. Appl. Math.}, 70(8):3342--3362, 2010.

\bibitem{NacTamTim11}
A.~Nachman, A.~Tamasan, and A.~Timonov.
\newblock Current density impedance imaging.
\newblock Preprint, 2011.

\bibitem{NasTam11}
M.~Z. Nashed and A.~Tamasan.
\newblock Structural stability in a minimization problem and applications to
  conductivity imaging.
\newblock {\em Inverse Probl. Imaging}, 5(1):219--236, 2011.

\bibitem{NegTonCon11}
M.~Negishi, T.~Tong, and R.~T. Constable.
\newblock Magnetic resonance driven electrical impedance tomography: a
  simulation study.
\newblock {\em IEEE Trans. Med. Imag.}, 30(5):828--837, 2011.

\bibitem{OzpIde05}
L.~{\"O}zparlak and Y.~Z. \.{I}der.
\newblock Induced current magnetic resonance-electrical impedance tomography.
\newblock {\em Physiol. Meas.}, 26:S289--S305, 2005.

\bibitem{Pel02}
L.~Pellerin.
\newblock Applications of electrical and electromagnetic methods for
  environmental and geotechnical investigations.
\newblock {\em Sur. Geophys.}, 23:101--132, 2002.

\bibitem{PeyYuLyoAlZFer96}
A.~J. Peyton, Z.~Z. Yu, G.~Lyon, S.~Al-Zeibak, J.~Ferreira, J.~Velez,
  F.~Linhares, A.~R. Borges, H.~L. Xiong, N.~H. Saunders, and M.~S. Beck.
\newblock An overview of electromagnetic inductance tomography: description of
  three different systems.
\newblock {\em Meas. Sci. Technol.}, 7:261--271, 1996.

\bibitem{PotWan09}
R.~Potthast and M.~Wannert.
\newblock Uniqueness of current reconstructions for magnetic tomography in
  multilayer devices.
\newblock {\em SIAM J. Appl. Math.}, 70(2):563--578, 2009.

\bibitem{RotWik98}
B.~J. Roth and J.~P. Wikswo.
\newblock Comments on ``{H}all effect imaging''.
\newblock {\em IEEE Trans. Biomed. Eng.}, 45(10):1294--1295, 1998.

\bibitem{SamCouTabBruRic05}
A.~Samou{\"e}lian, I.~Cousin, A.~Tabbagh, A.~Bruand, and G.~Richard.
\newblock Electrical resistivity survey in soil science: a review.
\newblock {\em Soil. Till. Res.}, 83:173--193, 2005.

\bibitem{Sch11}
O.~Scherzer, editor.
\newblock {\em Handbook of Mathematical Methods in Imaging}.
\newblock Springer, New York, 2011.

\bibitem{Schw63}
H.~P. Schwan.
\newblock Electric characteristics of tissues.
\newblock {\em Biophysik}, 1:198--208, 1963.

\bibitem{ScoJoyArmHen91}
G.~C. Scott, M.~L.~G. Joy, R.~L. Armstrong, and R.~M. Henkelman.
\newblock Measurement of nonuniform current density by magnetic resonance.
\newblock {\em IEEE Trans. Med. Imag.}, 10(3):362--374, 1991.

\bibitem{ScoJoyArmHen92}
G.~C. Scott, M.~L.~G. Joy, R.~L. Armstrong, and R.~M. Henkelman.
\newblock Sensitivity of magnetic-resonance current-density imaging.
\newblock {\em J. Magn. Reson.}, 97:235--254, 1992.

\bibitem{SeoWoo11}
J.~K. Seo and E.~J. Woo.
\newblock Magnetic resonance electrical impedance tomography ({MREIT}).
\newblock {\em SIAM Rev.}, 53(1):40--68, 2011.

\bibitem{Sli33}
L.~B. Slichter.
\newblock The interpretation of the resistivity prospecting method for
  horizontal structures.
\newblock {\em J. App. Phys.}, 4:307--322, 1933.

\bibitem{SonPauDehHar06}
N.~K. Soni, K.~D. Paulsen, H.~Dehghani, and A.~Hartov.
\newblock Finite element implementation of {M}axwell's equations for image
  reconstruction in electrical impedance tomography.
\newblock {\em IEEE Trans. Med. Imag.}, 25(1):55--61, 2006.

\bibitem{SteLeo07}
M.~Steffen and St. Leonhardt.
\newblock Development of the multichannel simultaneous magnetic induction
  measurement system {Musimitos}.
\newblock In H.~Scharfetter and R.~Merwa, editors, {\em 13th International
  Conference on Electrical Bioimpedance and 8th Conference on Electrical
  Impedance Tomography 2007}, IFMBE Proceedings Vol. 17, pages 448--451.
  Springer, 2007.

\bibitem{SurStuBarSwa88}
A.~J. Surowiec, S.~S. Stuchly, J.~R. Barr, and A.~Swarup.
\newblock Dielectric properties of breast carcinoma and the surrounding
  tissues.
\newblock {\em IEEE Trans. Biomed. Eng.}, 35:257--263, 1988.

\bibitem{SylUhl87}
J.~Sylvester and G.~Uhlmann.
\newblock A global uniqueness theorem for an inverse boundary value problem.
\newblock {\em Ann. Math.}, 125:153--69, 1987.

\bibitem{Tak97}
T.~Takagi et~al., editors.
\newblock {\em Electromagnetic nondestructive evaluation}.
\newblock IOS Press, 1997.

\bibitem{Uhl09}
G.~Uhlmann.
\newblock Electrical impedance tomography and {C}alder{\'o}n's problem.
\newblock {\em Inverse Probl.}, 25(12):123011, 2009.

\bibitem{WagZahGroPas04}
T.~A. Wagner, M.~Zahn, A.~J. Grodzinsky, and A.~Pascual-Leone.
\newblock Three-dimensional head model simulation of transcranial magnetic
  stimulation.
\newblock {\em IEEE Trans. Biomed. Eng.}, 51(9):1586--1598, 2004.

\bibitem{WanMaDeMNacJoy11}
D.~Wang, W.~Ma, T.~P. DeMonte, A.~I. Nachman, and M.~L.~G. Joy.
\newblock Radio-frequency current density imaging based on a 180$^\circ$ sample
  rotation with feasibility study of full current density vector
  reconstruction.
\newblock {\em IEEE Trans. Med. Imag.}, 30(2):327--337, 2011.

\bibitem{WenShaBal98}
H.~Wen, J.~Shah, and R.~S. Balaban.
\newblock Hall effect imaging.
\newblock {\em IEEE Trans. Biomed. Eng.}, 45(1):119--124, 1998.

\bibitem{Wid11}
T.~G. Widlak.
\newblock {Hybride Impedanztomographie}.
\newblock Master thesis, University of Vienna, Austria, Vienna, April 2011.

\bibitem{Wil10}
R.~A. Williams.
\newblock Landmarks in the application of electrical tomography in particle
  science and technology.
\newblock {\em Particuology}, 8:493--497, 2010.

\bibitem{WilKau81}
S.~J. Williamson and L.~Kaufman.
\newblock Biomagnetism.
\newblock {\em J. Magn. Magn. Mater.}, 22(2):129--201, 1981.

\bibitem{WooSeo08}
E.~J. Woo and J.~K. Seo.
\newblock Magnetic resonance electrical impedance tomography ({MREIT}) for high
  resolution conductivity imaging.
\newblock {\em Physiol. Meas.}, 29:R1--R26, 2008.

\bibitem{XuHe05}
Y.~Xu and B.~He.
\newblock Magnetoacoustic tomography with magnetic induction ({MAT-MI}).
\newblock {\em Phys. Med. Biol.}, 50:5175--5187, 2005.

\bibitem{ZhaWan04}
H.~Zhang and L.~Wang.
\newblock Acousto-electric tomography.
\newblock {\em Proc. SPIE}, 5320:145--149, 2004.

\bibitem{ZhaZhuHe10}
X.~Zhang, S.~Zhu, and B.~He.
\newblock Imaging electric properties of biological tissues by {RF} field
  mapping in {MRI}.
\newblock {\em IEEE Trans. Med. Imag.}, 29(2):474--481, 2010.

\bibitem{ZloRosAbb03}
S.~Zlochiver, M.~Rosenfeld, and S.~Abboud.
\newblock Induced-current electrical impedance tomography: a {2-D} theoretical
  simulation.
\newblock {\em IEEE Trans. Med. Imag.}, 22(12):1550--60, 2003.

\bibitem{ZouGuo03}
Y.~Zou and Z.~Guo.
\newblock A review of electrical impedance techniques for breast cancer
  detection.
\newblock {\em Med. Eng. Phys.}, 25:79--90, 2003.

\end{thebibliography}

\def\cprime{$'$} \providecommand{\noopsort}[1]{}

\end{document}